\documentclass[%
 amsmath,amssymb,
 aps,
prb,
twocolumn,
]{revtex4-1}

\usepackage{graphicx}
\usepackage{dcolumn}
\usepackage{bm}
\usepackage{bbm}
\usepackage{siunitx}
\usepackage{mathtools}
\usepackage{braket}
\usepackage{chemformula}
\usepackage{tabls}
\usepackage{verbatim}
\usepackage[toc,page]{appendix}
\usepackage{xspace}

\DeclareMathOperator{\Span}{Span}

\newcommand{\Rcal}{\mathcal{R}}
\newcommand{\Hcal}{{\hat{\mathcal{H}}}}

\newcommand{\Zcal}{\mathcal{Z}}

\newcommand{\bR}{\boldsymbol{R}}

\newcommand{\bq}{\boldsymbol{q}}

\newcommand{\bPhi}{\boldsymbol{\Phi}}
\newcommand{\bGamma}{\boldsymbol{\Gamma}}

\newcommand{\bUps}{\boldsymbol{\Upsilon}}

\newcommand{\bRcal}{\boldsymbol{\Rcal}}

\newcommand{\tpsi}{{\tilde\psi}}

\newcommand{\Tr}[1]{\textup{Tr}\left[#1\right]}
\newcommand{\Avg}[2]{\left\langle #1\right\rangle_{#2}}

\newcommand{\Avgquantum}[1]{\Avg{#1}{\hat\rho}}

\newcommand{\br}{{\bm r}}

\newcommand{\bJ}{\boldsymbol{J}}
\newcommand{\Qcal}{{\mathcal Q}}
\newcommand{\bQcal}{\boldsymbol{\Qcal}}
\newcommand{\Jcal}{{\mathcal J}}
\newcommand{\bJcal}{\boldsymbol{\Jcal}}

\newcommand{\secname}{{Sec.}}

\newcommand{\eqname}{{Eq.}}

\newcommand{\bpm}{\begin{pmatrix}}
\newcommand{\epm}{\end{pmatrix}}
\newcommand{\name}{NLSCHA\xspace}

\renewcommand{\figurename}{{Fig.}}

\begin{document}
	
	\preprint{APS/123-QED}
	
\title{A unified quantum framework for electrons and ions: \\The self-consistent harmonic approximation on a neural network curved manifold.}	
	
	\author{Lorenzo Monacelli$^{1,2}$, Antonio Siciliano$^2$, Nicola Marzari$^{1}$}
	
	\affiliation{$^1$ Theory and Simulation of Materials (THEOS), and \\ National Centre for Computational Design and Discovery of Novel Materials (MARVEL), École Polytechnique Fédérale de Lausanne (EPFL), Switzerland\\$^2$University of Rome ``Sapienza'', Italy}

	\date{\today}

\begin{abstract}
The numerical solution of the many-body problem of interacting electrons and ions is a key challenge in condensed matter physics, chemistry, and materials science. Traditional methods to solve the multi-component quantum Hamiltonian are usually specialized for one kind of particles -- electrons or ions -- and can suffer from a methodological gap when applied to the other ones. 

This work extends the self-consistent harmonic approximation, a proven successful technique for simulating quantum ions at finite temperatures in anharmonic crystals, to electrons. The approach minimizes the total free energy by optimizing an \emph{ansatz} density matrix, solving a fermionic self-consistent harmonic Hamiltonian on a curved manifold parametrized through a neural network. This approach preserves an analytical expression for entropy, enabling the direct computation of free energies and phase diagrams of materials.
By benchmarking this technique across several prototypical cases -- a double-well potential, the hydrogen atom, and the H$_2$ dissociation -- we demonstrate it can address both the ground- and excited-state properties of electronic systems, capture quantum tunneling and static electronic correlations, thereby providing a unified quantum framework of electrons and atomic nuclei. 
\end{abstract}
\maketitle


\section{Introduction}
Modeling materials from first principles requires the numerical solution of the fundamental equations of quantum mechanics for coupled electrons and ions. This is challenging if more than a few particles are involved. Thus, many approximations are typically employed to perform computational simulations and make predictions\cite{marzari_electronic-structure_2021}. The most common is treating electrons within a mean-field approach, like in density-functional theory (DFT)\cite{hohenberg_inhomogeneous_1964,burke_dft_2013}, simplifying the electron-electron interaction and significantly reducing complexity. Moreover, the quantum nature of atomic nuclei is usually neglected, as deviations from the classical behavior typically occur only at cryogenic temperatures. Finally, nuclei and electrons are mostly treated adiabatically, i.e., electrons remain in the ground state during the ionic dynamics. However, there are many materials in which these approximations fail, sometimes quite spectacularly. The mean-field approach to the electron-electron interaction is unsuitable in strongly correlated materials\cite{Georges1996}. Nuclear quantum effects play a significant role in the thermodynamic phase stability at high pressure or when the atomic mass is small, as is the case for hydrogen\cite{Monacelli2020_NatPhys,Monacelli2023_hydro}, hydrides\cite{Errea2016_H3S,Errea2020_Nature} and water\cite{Morrone2008, Cherubini2021,Ranieri2023,Cherubini2024}. The limits of computer simulations become even more severe when energy flows between electronic and nuclear excitations, leading to the breakdown of the adiabatic approximation.
This is relevant, e.g., when a system is near a conical intersection between electronic states, in metals with a low Fermi velocity\cite{Binci2021,Marchese2023,Girotto2023}, or in nonradiative electron-hole recombination\cite{Tong2022}. The complexity in modeling nonadiabatic phenomena hinders impactful technological advancements in materials design, like unconventional high-temperature superconductors, where low Fermi temperature coexists with a high superconductive critical temperature\cite{Uemura1989}, or preventing the nonradiative electron-hole recombination in metal-halide solar cells\cite{Tong2022}.

The most common approach to address nonadiabaticity is accounting for multiple electronic excited states as different potential energy landscapes (PES) for nuclei\cite{Tully1990}. The Ehrenfest dynamics is a mean-field approach to the electron-ion interactions, where ions move according to average forces on the different PES whose weights are evaluated by projecting the electronic dynamics within a time-dependent framework like TD-DFT. However, it has been shown that the Ehrenfest dynamic does not satisfy the detailed balance and leads to a wrong thermal state\cite{Nijjar2019}, thus being ineffective for simulating thermal equilibrium. The solution to this problem is achieved by the so-called ``surface hopping'' methods\cite{Tully1990}, in which each ionic trajectory evolves on a single PES with a probability of swapping electronic state during the dynamics\cite{Craig2005}. Final observables are then evaluated by averaging many different trajectories. However, these calculations are very expensive, requiring a dynamic treatment of multiple electronic excited states, and do not take into account intrinsically quantum nuclear effects\cite{Wang2016}, for which a path-integral reformulation is needed\cite{Shushkov2012} that further increases computational complexity. Moreover, these methods are explicitly devised for insulators or molecules. Their application is problematic in metallic systems where there is an infinite number of electronically excited states accessible by ionic excitations.

The opposite approach, which does not assume a multiple PES, is to treat electrons and nuclei within the same theory and solve the complete quantum problem. However, this poses substantial challenges, as methods traditionally used to tackle electrons and ions follow opposite strategies. While a mean-field approach is often enough for electrons, ions are intrinsically strongly correlated even in the simplest harmonic crystals; in fact, the lattice excited states, the phonons, are collective excitations where the ionic motion is correlated. Thus, one must choose methods capable of accounting for correlations, like Quantum Monte Carlo (QMC) or post-Hartre-Fock. However, these methods were devised to solve pure quantum states at $T=\SI{0}{\kelvin}$, as thermal excitations are not so relevant in electronic systems at room temperature; in fact, a temperature of $\SI{300}{\kelvin}$ correspond approximately to \SI{0.026}{\electronvolt}, which is usually small compared to the typical electronic excitation energies.
In contrast, temperature heavily affects ions. The thermal excitations of nuclei trigger most of the physically relevant phenomena for materials, like phase transitions and thermal expansion. Path-integral molecular dynamics (PIMD) is the state-of-art for nuclei when quantum effects are relevant\cite{Ceperley_path_1995}. 
The computational complexity required to converge PIMD calculations depends on the temperature, diverging for $T\to\SI{0}{\kelvin}$. This makes PIMD suitable for studying high-temperature states near the limit where particles behave classically, and it is ideal for nuclei. However, electrons at room temperature are very far from the classical limit. For these reasons, path-integral Monte Carlo simulations have been applied to the electrons-ion plasma far above room temperature \cite{Ceperley_path_1995,Driver2012}, in the range of $10^4-10^9$~K. 
Moreover, all the aforementioned methods suffer from the complexity arising when computing the phase diagram of materials. No direct computable expression for the entropy exists in these methods, and differences between the free energy of the structures need to be evaluated through thermodynamic integration, which is often out of reach even when fast potentials are available within the BO approximation. 

In this work, we introduce a new framework that can be applied simultaneously to both electrons and ions at room temperature. To this aim, we extend the self-consistent harmonic approximation\cite{Errea2014,Monacelli2021,miotto_fast_2024} (SCHA), a technique devised initially to solve the adiabatic quantum nuclear problem, to electrons. The SCHA was successfully employed to study many equilibrium properties of matter where quantum atomic effects are dominant, like the phase diagram of high-pressure hydrogen\cite{Borinaga2016mol,Borinaga2016atom, Monacelli2020_NatPhys,Monacelli2023_hydro}, hydrides superconductors\cite{Errea2016_H3S,Errea2020_Nature}, hydrate clathrates\cite{Ranieri2023}, metal-halide perovskite solar cells\cite{Monacelli2023}, and the emergency of charge-density waves in 2D and bulk transition metal dichalcogenides\cite{SkyZhou2020,Bianco2019NbS2,Bianco2020_NbSe2,Diego2021}.
The original method approximates the quantum density matrix of nuclei as the Gaussian equilibrium state of a trial quantum harmonic Hamiltonian. Gaussian density matrices provide an appropriate description for nuclei in a crystal as they oscillate around fixed equilibrium positions. However, they are not suited to describe electronic wavefunctions, as the lighter mass of electrons produces delocalized states where deviations from the Gaussian shape become much more relevant, leading to the breakdown of the SCHA in electronic systems. Besides, while the indistinguishability of nuclei plays a negligible part in thermodynamics, electronic exchange gives rise to the Pauli exclusion principle, which plays a dominant role in atomic physics.

\begin{figure*}[bhtp]
    \centering
    \includegraphics[width=\linewidth]{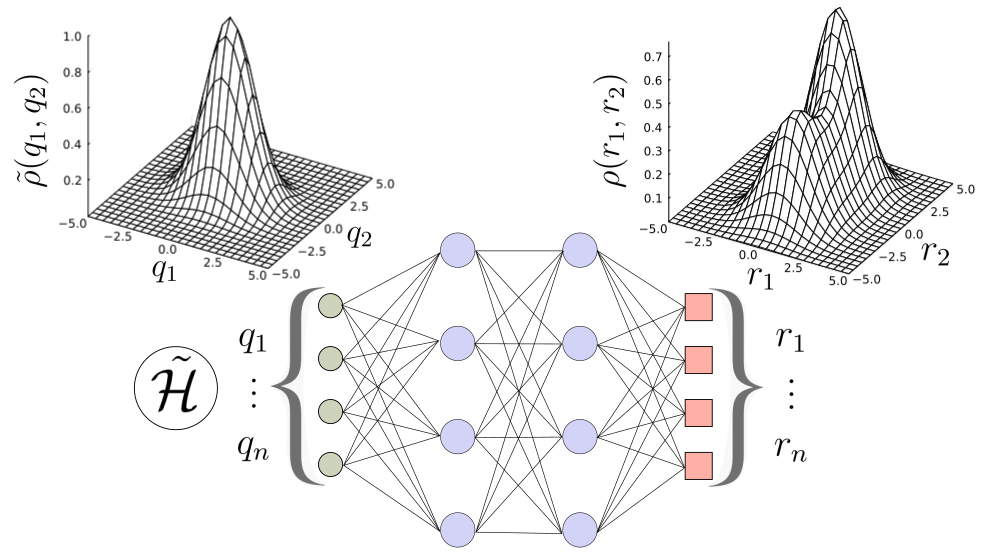}
    \caption{Schematic representation of the SCHA on a neural network curved manifold (nonlinear SCHA: NLSCHA). The Gaussian density matrix $\tilde\rho$, solution of the trial harmonic Hamiltonian $\tilde {\mathcal H}$ in an auxiliary space $\bq$ (the curved manifold) is transformed into real space thanks to the transformation $\br(\bq)$ defined through a neural network. The nonlinearity of the transformation deforms the Gaussian shape and introduces new features on the final density matrix, including correlations beyond the original distribution.}
    \label{fig:method}
\end{figure*}

Here, we extend the SCHA framework to handle indistinguishable particles with spin. Our novel approach also enhances the density matrix beyond the Gaussian approximation, but it keeps a computational cost comparable to the original SCHA. This is achieved thanks to the recently developed Nonlinear SCHA\cite{siciliano_beyond_2024,siciliano_beyond_2024-1}, where the degrees of freedom of the trial density matrix are augmented by introducing a curved manifold on which the Gaussian density matrix is defined. The curved manifold deforms the density matrix, increasing the variational space available beyond Gaussians for optimizing the density matrix to minimize the free energy. To parameterize and optimize the curved manifold, we introduce a neural network representation, where the systematic addition of hidden layers further improves the variational space of the density matrix.
The core of the approach is illustrated in \figurename~\ref{fig:method}, where the original Gaussian density matrix is deformed through the transformation encoded by the neural network.

We start by briefly reviewing the SCHA in \secname~\ref{sec:scha}, extending the theory in the case of indistinguishable particles, with particular care to the case of fermions (\secname~\ref{sec:harm:fermions}). Then, we introduce the curved manifold that allows the wavefunction to deviate from the original Gaussian shape (\secname~\ref{sec:nonlinear}). We discuss the neural network parametrization of the curved manifold (\secname~\ref{sec:neural:network}) and how to encode crystal symmetries within it. \secname~\ref{sec:electrons} discusses the specific requirements of the neural network parametrization to deal with fermionic degrees of freedom.
In \secname~\ref{sec:applications}, we apply the new theory to study several challenging cases where the SCHA fails, such as the profound double-well potential (to model quantum tunneling), the hydrogen atom (to model the electron-ion cusp), and the \ch{H2} dissociation (to model electron-electron correlations), both in the ground bonding state (singlet) and in the excited antibonding one (triplet). We then address limitations and future perspectives of this approach in \secname~\ref{sec:limits}.

\section{Self-Consistent Harmonic Approximation}
\label{sec:scha}

The self-consistent harmonic approximation is a variational approach to the solution of the time-independent Schr\"odinger equation:
\begin{equation}
    \hat H\ket \psi = E \ket \psi ,
\end{equation}
where $\ket{\psi}$ is the ground-state wave-function of a generic Hamiltonian operator $\hat H$, and $E$ is its associated energy. We indicate an operator in the Hilbert space with a hat.
The SCHA was successfully employed to solve the nuclear quantum problem, where $\hat H$ depends only on atomic coordinates, and the electrons' degrees of freedom are integrated out within the Born-Oppenheimer approximation. Here, we extend the formalism to electrons. Thus, $\hat H$ depends on the position operators $\hat R_i$ of both nuclei and electrons, overcoming the Born-Oppenheimer approximation. We neglect the role of spin interactions, such as spin-orbit coupling, and assume that the total spin operator $\hat {S^2}$ and its projection on one axis $\hat{S_z}$ commute with the Hamiltonian $\hat H$.
The SCHA constrains the wavefunction $\ket\tpsi$ to the ground state of an auxiliary harmonic Hamiltonian $\Hcal[\bRcal, \bPhi]$ depending on the parameters $\bRcal$ and $\bPhi$ as
\begin{equation}
    \Hcal[\bRcal, \bPhi] = \hat K + \frac 12 \sum_{ij} (\hat R_i - \Rcal_i)\Phi_{ij}(\hat R_j - \Rcal_j),
    \label{eq:H:aux:psi}
\end{equation}
\begin{equation}
    \Hcal[\bRcal, \bPhi] \ket{\tpsi[\bRcal,\bPhi]} = {\mathcal E} \ket{\tpsi[\bRcal,\bPhi]},
\end{equation}
where $\hat K$ is the kinetic energy operator, $\hat R_i$ is the position operator of the $i$-th particle (electron or nucleus). Latin indices indicate both the particle and the Cartesian component, and bold symbols represent tensorial quantities (e.g., vectors and matrices). The vector $\Rcal_i$ indicates the average position of particle $i$ (centroids), while $\bPhi$ are auxiliary force constant matrices encoding the quantum fluctuations and the correlations between different particles. These parameters must minimize the total energy within the Rayleigh-Ritz variational principle 
\begin{equation}
    E^\text{(SCHA)} = \min_{\bRcal, \bPhi} \braket{\tpsi[\bRcal,\bPhi]| \hat H | \tpsi[\bRcal,\bPhi]}.
    \label{eq:rayleigh-ritz}
\end{equation}
The minimum of \eqname~\eqref{eq:rayleigh-ritz} is satisfied by the self-consistent equations\cite{SSCHA}
\begin{equation}
    \braket{\tpsi[\bRcal,\bPhi]|\frac{d\hat V}{dR_i}|\tpsi[\bRcal,\bPhi]} = 0,
    \label{eq:self-consistent:cent}
\end{equation}
\begin{equation}
    \braket{\tpsi[\bRcal,\bPhi]|\frac{d^2\hat V}{dR_idR_j}|\tpsi[\bRcal,\bPhi]} = \Phi_{ij},
    \label{eq:self-consistent:psi}
\end{equation}
where $\hat V = V(\hat \bR)$ is the interaction potential as a function of the position operators of the particles.
By solving \eqname~\eqref{eq:self-consistent:cent} and \eqname~\eqref{eq:self-consistent:psi} $\bRcal$ and $\bPhi$ are updated to get a new auxiliary Hamiltonian $\Hcal[\bRcal^\text{(next)},\bPhi^\text{(next)}]$, defining a ground state $\ket{\tpsi^\text{(next)}}$ for the next iteration. The process is repeated until convergence of \eqname~\eqref{eq:self-consistent:cent} and \eqname~\eqref{eq:self-consistent:psi}.

The SCHA works at finite temperature by replacing the wavefunction with the density matrix $\hat\rho$ and the averages as traces. In this case, the auxiliary Hamiltonian defines the density matrix through the usual equilibrium relationship
\begin{equation}
	\hat \rho[\bRcal, \bPhi] = \frac{\exp\left( - \beta \Hcal[\bRcal, \bPhi]\right)}{\Zcal},
	\label{eq:rho:scha}
\end{equation}
\begin{equation}
	{\Zcal}= \Tr{\exp\left( - \beta \Hcal[\bRcal, \bPhi]\right)}.
	\label{eq:rho:scha}
\end{equation}
We omit the explicit dependence of $\hat\rho$ from the $\bRcal$ and $\bPhi$ in the following equations. 
Indeed, the equilibrium solution is obtained by minimizing the free energy functional $\mathcal F[\hat \rho]$\cite{SSCHA}
\begin{equation}
    \mathcal F[\hat \rho] = \Avgquantum{\hat H} - T S\left[\hat\rho\right],
    \label{eq:free:energy}
\end{equation}
where $T$ is the temperature and $S[\hat\rho]$ is the entropy functional, defined as
\begin{equation}
    S[\hat \rho] = -k_b\Avgquantum{\ln\hat\rho},
\end{equation}
where the quantum averages are
\begin{equation}
    \Avgquantum{\hat H} = \Tr{\hat\rho \hat H},
\end{equation}
and $k_b$ is the Boltzmann constant.
The equilibrium free energy $F$ is always lower than the free energy functional $\mathcal F[\hat\rho]$ for any $\hat \rho$ that is not the exact equilibrium one. Thus, the self-consistent equations to update $\bRcal$ and $\bPhi$ are obtained by minimizing \eqname~\eqref{eq:free:energy}
\begin{equation}
    \Avgquantum{\frac{dV}{dR_i}} = \Tr{\hat \rho \frac{dV}{dR_i}} =  0,
\end{equation}
\begin{equation}
    \Avgquantum{\frac{d^2V}{dR_idR_j}} = \Tr{\hat \rho \frac{d^2V}{dR_idR_j}} = \Phi_{ij}[\hat\rho].
    \label{eq:self-consistent:rho}
\end{equation}

\section{SCHA for electrons}
\label{sec:harm:fermions}
In this section, we apply the SCHA to electrons, showing how it is possible to deal with a system of indistinguishable fermions within the SCHA framework.
To describe a fermionic (or bosonic) system, the auxiliary harmonic Hamiltonian must commute with the operator $\bm S_{ij}$ that swaps the position of particle $i$ with $j$, for any $i$ and $j$.
This imposes the constraints on $\bRcal$ and $\bPhi$
\begin{equation}
\forall i,j \qquad
    \bm S_{ij}\bRcal = \bRcal ,
    \qquad
    \bm S_{ij}\bPhi \bm S_{ij}^\dagger = \bPhi.
\end{equation}
The result is that the centroid position must be the same for all electrons, reducing its independent components from $3N$ to $3$. The only surviving elements of the $\bPhi$ are the 6 independent degrees of freedom that encode the coupling between an electron with itself ($\bPhi^\text{(same)}$) and the 9 degrees of freedom that couple two different electrons ($\bPhi^\text{(diff)}$), for a total of 15 degrees of freedom (versus $3N(3N+1)/2$ for distinguishable particles)
\begin{align}
    \Hcal = \hat K& + \frac 12\sum_{i\alpha\beta}(\hat R_{i\alpha} - \Rcal_\alpha)\Phi^{\text{(same)}}_{\alpha\beta}(\hat R_{i\beta} - \Rcal_\beta) +\nonumber \\
    & + \frac 12\sum_{\substack{i\neq j\\\alpha\beta}}(\hat R_{i\alpha} - \Rcal_\alpha)\Phi^{\text{(diff)}}_{\alpha\beta}(\hat R_{j\beta} - \Rcal_\beta).
    \label{eq:scha:electrons}
\end{align}
In this section, we explicitly encode with $i\alpha$ and $j\beta$ the index of the particle and the cartesian coordinate. While $\bPhi^\text{(same)}$ is hermitian, the same is not necessarily valid for $\bPhi^\text{(diff)}$. Notably, $\bPhi^\text{(diff)}$ is a coupling term between electrons. Therefore, the SCHA differs from other mean-field approaches like Hartree-Fock (HF) and density-functional theory (DFT) because the auxiliary Hamiltonian already accounts for a certain degree of electronic correlation. \eqname~\eqref{eq:scha:electrons} can be
diagonalized exactly, and the resulting eigenstates cannot be separated into products of independent particle wave functions. However, it is challenging to disentangle the bosonic-fermionic statistics of the states. Also, we prove in \appendixname~\ref{app:harmonic:fermions} that $\bPhi^\text{(diff)}\to 0$ in the $N\to\infty$ limit.
For these reasons, we further restrict the problem to noninteracting fermions in a harmonic potential, neglecting the correlations in the harmonic Hamiltonian. We show in \secname~\ref{sec:neural:network} that this correlation can be restored on the density matrix \emph{a posteriori} with a linear deformation of the curved manifold.

\begin{figure}[hbtp]
\centering
\includegraphics[width=\columnwidth]{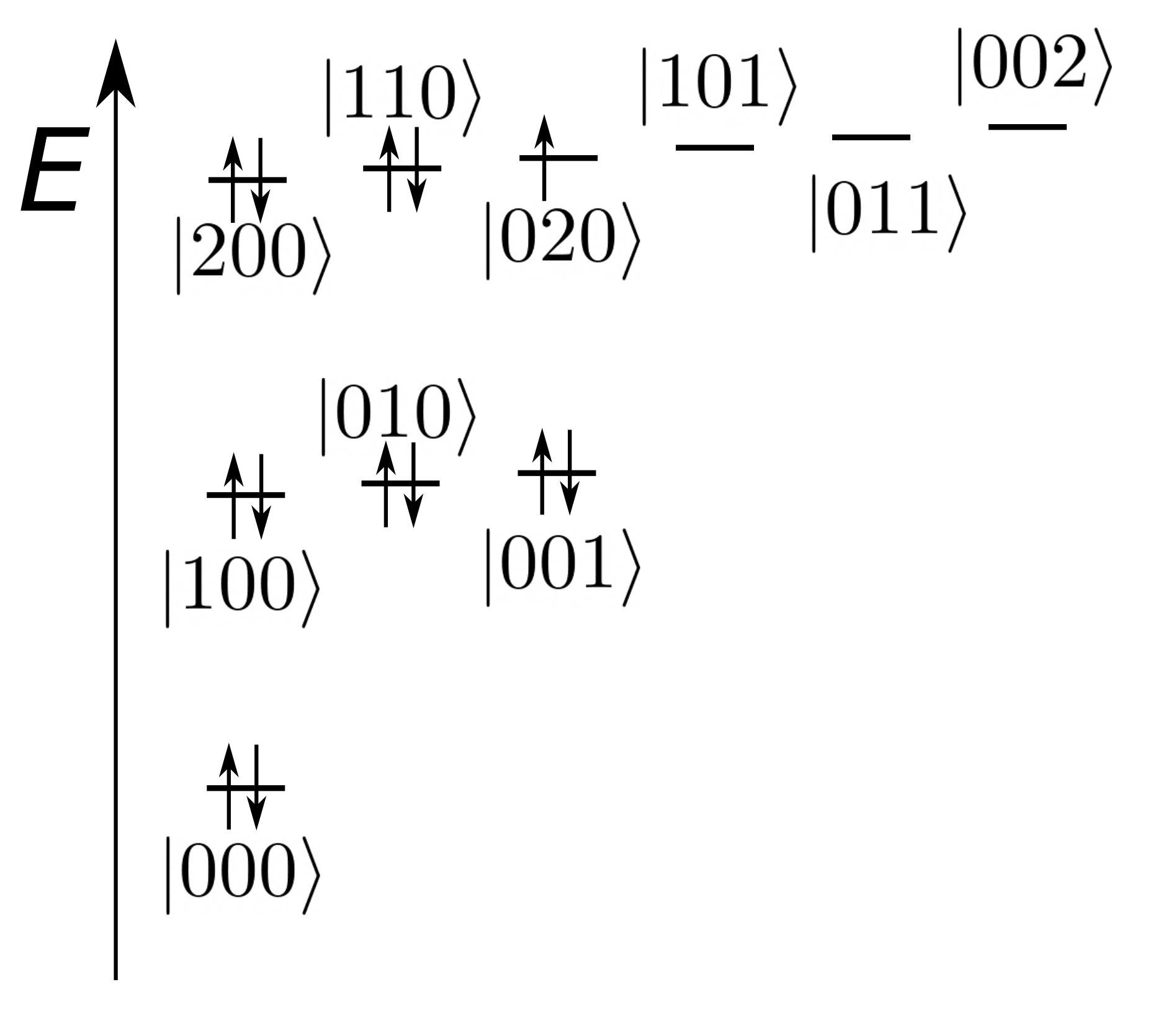}
\caption{
\small States of the noninteracting fermionic oscillator in 3D. Electrons populate the excited states of the harmonic Hamiltonian $\ket {n_1n_2 n_3}$ according to the Fermi-Dirac statistics and the spin multiplicity.\label{fig:harmonic:fermi}}
\end{figure}

Without $\bPhi^\text{(diff)}$, the Hamiltonian \eqname~\eqref{eq:scha:electrons} is noninteracting, and it can be diagonalized by identifying the three eigenvectors of $\bPhi^{\text{(same)}}$. Introducing the standard creation $a_\mu^\dagger$ and annihilation $a_\mu$ operators of the harmonic oscillator along each of these three modes $\mu$, we have
\begin{equation}
    \Hcal = \frac 12 \sum_{\mu = 1}^3 \omega_\mu \left(2a_\mu^\dagger a_\mu + 1\right).
    \label{eq:quantum:harmonic}
\end{equation}
A 3-state vector $\ket{n_1 n_2 n_3}$ uniquely identifies the state, encoding the excited level on each polarization mode. The overall fermionic wavefunction is obtained by constructing a Slater determinant with these orbitals similar to standard mean-field approaches, constraining the rules for populating states given by the Pauli exclusion principle, as shown in \figurename~\ref{fig:harmonic:fermi}.
Notably, the electronic oscillator's ground state is a Gaussian only up to 2 electrons. Excited orbitals are occupied for more than two electrons, and the many-body ground state acquires the nodes typical of fermionic wavefunctions. In the variational frameworks of the SCHA, the nodes are not fixed and depend on the parameters $\bRcal$ e $\bPhi^\text{(same)}$. Thus, the electronic SCHA does not rely on any fixed-node approximation necessary for applying path-integral Monte Carlo or diffusion Monte Carlo to fermions\cite{foulkes_quantum_2001}. 

\section{SCHA in a curved manifold}
\label{sec:nonlinear}

While the SCHA has been shown to work exceptionally well in investigating the anharmonic quantum motion of nuclei in crystals\cite{Errea2016_H3S,Aseginolaza2019,verdi_quantum_2023,Romanin2021,ranalli_temperature-dependent_2023,Pedrielli2022,Monacelli2020_NatPhys,Monacelli2023_hydro,Errea2020_Nature}, its limitations arise from constraining the density matrix to solve a harmonic Hamiltonian. In the case of distinguishable particles like atomic nuclei, the real-space expression $\rho(\bR, \bR') = \braket{\bR'|\hat\rho|\bR}$ of the SCHA density matrix is a Gaussian\cite{Monacelli2021}. A Gaussian distribution fits the fluctuations of nuclei vibrating in a solid lattice. Still, it fails to capture ionic diffusion, molecular rotations, quantum tunneling, or cases where the wave function extends beyond the interatomic distance. These cases are significant for electrons, where their lighter mass by at least 3 orders of magnitude causes a quantum delocalization across a broad region of space, often encompassing many atoms by forming chemical bonds.

Overcoming the SCHA has proven more difficult than expected: any attempt to augment the auxiliary harmonic Hamiltonian with extra anharmonic terms prevents the analytical solvability of the auxiliary system. This results in the loss of many of the technique's advantages, among which is the analytical expression for the entropy\cite{SSCHA}, fundamental to computing the phase diagram of materials.
An alternative strategy consists of maintaining the auxiliary Hamiltonian harmonic, introducing a nonlinear change of variables between the Cartesian positions $\bR$ and the degrees of freedom of the auxiliary Hamiltonian $\bq$\cite{siciliano_beyond_2024,siciliano_beyond_2024-1}.
This defines the auxiliary harmonic Hamiltonian on a manifold whose curvature is parametrized by the invertible transformation $\bR(\bq)$. This approach is similar to what was proposed by Gigy and Hamann\cite{Gygi1992,Hamann1995} in the context of density functional theory to improve the basis set variationally. In this case, however, we utilize the nonlinear change of variable to capture both deviations from Gaussianity and many-body correlation effects.
The expression of physical observables must be the same on the curved manifold $\bq$ and real space $\bR$.
\begin{equation}
    \Avgquantum{\hat O} = \Tr{\hat{\tilde \rho} \hat O}
\end{equation}
\begin{equation}
    \Avgquantum{\hat O} = \int d\bR d\bR' O(\bR, \bR')\rho(\bR', \bR) ,
\end{equation}
\begin{equation}
    \Avgquantum{\hat O} =\int d\bq d\bq' \tilde O(\bq, \bq')\tilde\rho(\bq', \bq),
\end{equation}
where we indicated with a tilde $\tilde\cdot$ quantities evaluated on the curved manifold.
By imposing that the density matrix conserves the probability density in the position space, 
we can prove that diagonal observables like $O(\bR, \bR') \propto \delta(\bR - \bR')$ transform like
\begin{equation}
    \tilde O(\hat\bq) = O(\bR(\hat \bq)),
    \label{eq:transform:diagonal}
\end{equation}
and the density matrix as
\begin{equation}
    \tilde\rho(\bq, \bq') =\rho(\bR(\bq), \bR(\bq')) \sqrt{\det\bJcal(\bR(\bq)) \det\bJcal(\bR(\bq'))} ,
    \label{eq:transform:rho}
\end{equation}
where $\bJcal$ is the jacobian of the transformation
\begin{equation}
    \Jcal_{ab}(\bq) = \frac{\partial R_a}{\partial q_b}.
\end{equation}
The transformation of \eqname~\eqref{eq:transform:rho} preserves the probability density in the position space. A more complex transformation for the observable needs to be derived if $O(\bR, \bR')$ is not diagonal in the position basis, e.g., for the kinetic energy.

We have, thus, two different density matrices: $\tilde\rho(\bm q, \bm q')$, defined on the curved manifold and solution of the auxiliary harmonic Hamiltonian (\eqname~\ref{eq:rho:scha}), 
and $\rho(\bR, \bR')$, the transformed density matrix in Cartesian space.
Notably, $\rho(\bR, \bR')$ is the real density matrix of our solution, which depends on the parameters that define the auxiliary Hamiltonian ($\bRcal$ and $\bPhi$) and on the curvature of the manifold, defined by the metric tensor $g_{ab}$ as
\begin{equation}
    g_{ab}(\bq) = \sum_c \frac{\partial R_c}{\partial q_a}\frac{\partial R_c}{\partial q_b} = \sum_c \Jcal_{ca}\Jcal_{cb},
\end{equation}
\begin{equation}
    \sqrt{\det {\bm g}} = \sqrt{\det\bJcal(\bR(\bq)) \det\bJcal(\bR(\bq'))} .
\end{equation}
This allows us to increase the degrees of freedom of the ansatz density matrix thanks to the metric tensor $\bm g$ while keeping the auxiliary system harmonic. 
This approach maintains all the advantages of the original SCHA. In fact, all observables can be evaluated in the curved manifold, where the density matrix is harmonic. For convenience, we name the method nonlinear SCHA (or \name) throughout the rest of the manuscript. 

\subsection{Free energy in the curved manifold}
\label{sec:free:energy}
The target of the SCHA as a variational theory is to optimize the trial density matrix $\rho(\bR, \bR')$ to minimize the free energy.
The free energy $F[\hat \rho]$ is a functional of the density matrix comprising the internal energy $U[\hat\rho] = \Avgquantum{\hat H}$ and the entropy $S[\hat\rho]$.
The most complex part of the free energy is the entropy, for which the expression is challenging to be converged numerically
\begin{equation}
    S[\hat \rho] = - k_b \braket{\ln\hat\rho}.
    \label{eq:entropy:general}
\end{equation}
In fact, due to the logarithm, even very high energy states are important in the average of \eqname~\eqref{eq:entropy:general}, and it cannot be computed as an average on the sampled configurations.
However, if the nonlinear transformation $\bm R(\bm q)$ is bijective on $\mathbb{R}^{3N}\to\mathbb{R}^{3N}$, the entropy of $\rho(\bR, \bR')$ coincides with the one of $\tilde\rho(\bq, \bq')$\cite{siciliano_beyond_2024}.
Thus, $S[\hat \rho]$ can be evaluated conveniently on the curved manifold, where the density matrix solves the auxiliary harmonic Hamiltonian $\Hcal[\bRcal, \bPhi]$ for which the energy levels are known analytically. 
In the case of distinguishable particles, like atomic nuclei, the entropy has a simple expression depending on the frequencies $\omega_\mu$ of the auxiliary Hamiltonian\cite{ashcroft_solid_1976}
\begin{equation}
    \sum_{b}\frac{\Phi_{ab}}{\sqrt{m_am_b}} e_\mu^a = \omega_\mu^2 e_\mu^b,
\end{equation}
\begin{equation}
S[\rho(\bm R, \bm R')] = \frac{k_b}{2}\sum_\mu\left[ \frac{\beta \hbar\omega_\mu}{\tanh\frac{\beta \hbar\omega_\mu}{2}} - 2 \log \left(\sinh\frac{\beta \hbar\omega_\mu}{2}\right)\right] ,
\label{eq:S:phonons}
\end{equation}
where $m_a$ is the mass of the $a$ particle, $\beta = (k_bT)^{-1}$ is the Boltzmann factor. In the more general case of mixtures between distinguishable and indistinguishable particles, the energy levels of the auxiliary harmonic Hamiltonian must be appropriately populated, accounting for the specific Fermi-Dirac or Bose-Einstein statistic and the particles' spin. Thankfully, the auxiliary harmonic Hamiltonian can be decoupled between each different kind of particle, separating the indistinguishable fermions (electrons) and the distinguishable ions. The electronic entropy is thus evaluated in a noninteracting harmonic system, as discussed in \secname~\ref{sec:harm:fermions}. Let $f_i$ be the occupation number of the $i$ state of the auxiliary harmonic Hamiltonian, the resulting electronic entropy is
\begin{equation}
    S[\hat\rho] = -k_b\sum_i\left[f_i\ln(f_i) + (1 - f_i)\ln(1 - f_i)\right].
\label{eq:S:electrons}
\end{equation}

The internal energy $U[\hat\rho]$ accounts for kinetic and potential energy. The potential energy is a diagonal operator in the position of the particles $\hat R$, and can be evaluated on the curved manifold following \eqname~\eqref{eq:transform:diagonal}
\begin{equation}
\Avgquantum{\hat V} = \int d\bq \; V\left(\bR(\bq)\right) \tilde \rho(\bq).
\label{eq:potential}
\end{equation}
\eqname~\eqref{eq:potential} is the average interacting potential in the curved manifold. The practical computation of $\Avgquantum{\hat V}$ through Monte Carlo consists of generating a random ensemble of particles on the positions $\bq$ in the manifold exploiting the Gaussian shape of the auxiliary probability distribution given by $\tilde \rho(\bq)$, transforming the coordinates from the manifold $\bq$ to Cartesian space $\bR$ with the transformation $\bR(\bq)$, and averaging the values of the potential $V(\bR)$ in Cartesian space. 
Compared with other approaches like PIMD or QMC, the generation of the ensemble is faster and does not require any Metropolis algorithm or thermalization process, as it occurs on a Gaussian distribution in the curved manifold.

The kinetic energy is more complex as it is nondiagonal in position space. Its expression $\tilde K(\bq, \bq')$ on the curved manifold is derived by transforming the integration variables from Cartesian space
\begin{equation}
\Avgquantum{K} = -\sum_i\frac{\hbar^2}{2m_i} \int d\bR d\bR' \delta(\bR - \bR') \frac{\partial^2}{\partial R'_i} \rho(\bR, \bR')\label{eq:kin:classic}
\end{equation}
In \appendixname~\ref{app:kinetic} we derive the explicit expression for the kinetic energy in the curved manifold. In practice, it can be computed as
\begin{align}
    \Avgquantum{K} = \sum_i \frac{1}{2m_i}&\bigg[\sum_{ab}\int d\bm q d\bm q' \gamma^{ab}_i(\bm q) \frac{\partial^2\tilde\rho(\bm q, \bm q')}{dq_adq'_b}\delta(\bm q - \bm q') + \nonumber \\
    & +\frac 12\sum_a \left<\beta_i^a(\bm q)\frac{d\ln\tilde\rho(\bm q)}{dq_a}\right>_{\tilde\rho} + \left<\alpha_i(\bm q)\right>_{\tilde\rho}\bigg]
    \label{eq:kin}
\end{align}
Here, $\tilde \rho(\bm q)$ is the probability density in the auxiliary flat space: the diagonal elements in real space of the density matrix, i.e. $\tilde\rho(\bm q) \dot = \tilde\rho(\bm q, \bm q)$. $\left<f(\bm q)\right>_{\tilde\rho}$ indicate an expected value of the diagonal operator $\hat f$ in the positions $\hat \bq$ on the auxiliary flat space, and $\bm \alpha(\bm q)$, $\bm\beta(\bm q)$ and $\bm\gamma(\bm q)$ are operators that depend only the local curvature of the manifold (the nonlinear transformation). Their explicit expression is reported in \appendixname~\ref{app:kinetic}. 
Indeed, in the case of a flat manifold, $\bm \alpha = \bm \beta = 0$, $\gamma_i^{ab}(\bq) = \delta_{ai}\delta_{bi}$, and we recover the kinetic energy of a flat space (\eqname~\ref{eq:kin:classic}). The complete kinetic energy on the curved manifold comprises a term (the first one) which represent the kinetic energy of the auxiliary flat space, where the masses are contracted with a vector that depends on the metric tensor: $\gamma_i^{ab}(\bq)$. This can be interpreted as a change in the mass of the particles due to the nonlinear transformations (see \appendixname~\ref{app:mass} for more details). Besides this term, the kinetic energy also includes a correction that is diagonal in position space, thus behaving like an effective potential energy. This extra term encodes the correlations introduced by the curved manifold, similarly to the exchange-correlation potential defined within DFT.
Notably, \eqname~\eqref{eq:kin} differs from the equation presented in Ref.\cite{siciliano_beyond_2024}, as here we do not perform any hypothesis on the Gaussian shape of the auxiliary density matrix $\tilde\rho(\bq, \bq')$. This way, \eqname~\eqref{eq:kin} can also be applied to electrons, excited states, or any nonharmonic auxiliary Hamiltonian. 

\section{Neural network parametrization of the curved manifold}
\label{sec:neural:network}

In \secname~\ref{sec:free:energy}, we derived the free energy expression on the curved manifold, where the density matrix is the ground state of an auxiliary harmonic Hamiltonian, and the free energy can be computed efficiently. 
To solve the quantum problem beyond the SCHA, we need to minimize the free energy by optimizing the parameters defining the auxiliary Hamiltonian and the manifold.
The auxiliary Hamiltonian depends on the centroids $\bRcal$ and the effective force constants $\bPhi$. Ref.~\cite{siciliano_beyond_2024-1} devised a specific manifold from a cartographic representation of spherical coordinates to describe molecular rotations. While it provides an excellent solution to enhance the SCHA by enabling molecules to rotate thanks to the manifold's curvature, it is too restrictive to capture the complex features of the electronic wavefunction.
Here, we introduce a general transformation that can be systematically improved to parameterize any possible curved manifold, thus significantly enhancing the variational space of density matrices.

The manifold is defined through the transformation $\bR(\bq)$. This function maps $\mathbb R^N\to\mathbb R^N$, is bijective, continuous, and differentiable. We define the
Gaussian stretchers as the class of manifolds parametrized by the transformation
\begin{equation}
    \bR(\bq) =\bQcal + \tilde \bq \left[1 + (e^\zeta - 1) \exp\left(-\frac 12 \sum_{cd} \tilde q_c \Gamma_{cd} \tilde q_d\right)\right] ,
    \label{eq:gauss:stretch}
\end{equation}
where
\begin{equation}
    \tilde \bq = \bq - \bQcal ,
\end{equation}
and $\bGamma$ is positive definite.
As the name suggests, this transformation stretches the space, deforming the density matrix in the position identified by $\bQcal$ along the principal axis of the $\bGamma^{-1}$ covariance matrix, with intensity $\zeta$.
Outside the central position $\bQcal$ and the range of the covariance matrix $\bGamma^{-1}$, the exponential of \eqname~\eqref{eq:gauss:stretch} becomes small and the manifold flat ($\bR = \bq$).
Therefore, the Gaussian stretcher only warps the space inside the Gaussian by stretching the density matrix with a positive curvature if $\zeta > 0$ (reducing the probability distribution at the center and accumulating it on the edges) or compressing it with a negative curvature if $\zeta < 0$ (reducing the probability on the edges to accumulate it on the center).
The parameterization of the Gaussian stretcher is only valid for $\zeta < \zeta_c \approx 1.1$, above which the transformation defined in \eqname~\eqref{eq:gauss:stretch} is no longer bijective.

Each Gaussian stretcher has the same number of degrees of freedom as the original SCHA algorithm plus $\zeta$ (the $\bGamma$ symmetric 2-rank tensor is equivalent to the auxiliary force constant matrix $\bPhi$, and the $\bQcal$ vector has the same length as the centroids $\bRcal$). 
Moreover, multiple Gaussian stretchers can be stacked together in layers as in neural networks:
\begin{equation}
    \bq_{n+1} = \bQcal_{n} + \tilde\bq_n\left[1 + (e^{\zeta_n} - 1)\exp\left(-\frac 12 \sum_{cd}\tilde {q_n}_{c} \Gamma^{(n)}_{cd} \tilde {q_n}_d\right)\right],
    \label{eq:activ:func}
\end{equation}
\begin{equation}
    \bq_0 = \bR \qquad \bQcal_0 = \bRcal.
\end{equation}
Each new layer introduces three new parameters: $\zeta_n$, $\bQcal_n$, and $\bGamma^{(n)}$. In the case of concatenated Gaussian stretchers, the resulting manifold exhibits a complex curvature that is fine-tuned in any position by a layer with a nearby centroid $\bQcal_n$.
The transformation resembles a neural network where the activation function for each neuron is given by \eqname~\eqref{eq:gauss:stretch}. In \appendixname~\ref{app:gauss:combined}, we evaluate the metric tensor and the expression of the kinetic energy. The manifold parameters are optimized through a standard ADAM algorithm\cite{ADAM}, where the gradient of the quantum free energy is evaluated with back-propagation, a technique commonly employed in neural networks.

Equivariance under specific symmetries can be enforced on the Gaussian stretcher manifold. This is pivotal for ensuring compliance with fundamental laws of physics, like the acoustic sum rule (equivariance with respect to global translations) and the permutation of indistinguishable particles. Moreover, this characteristic extends to any symmetry operation commuting with the original Hamiltonian as the crystallographic symmetry group.
In particular, let $\bm S$ be the symmetry operator;
the manifold is equivariant under $\bm S$ if 
\begin{equation}
    \bR(\bm S \bq) = \bm S \bR(\bq).
\end{equation}
This condition can be applied layer by layer on the Gaussian stretcher as
\begin{equation}
    \bm S \bm {q}_{n+1} = \bQcal_n  + (\bm S \bm {q}_{n} - \bQcal_n)f_n(\bm S\bm {q}_n),
    \label{eq:simp}
\end{equation}
where
\begin{equation}
    f_n(\bq) = 1 + \left(e^{\zeta_n} - 1\right)
    \exp\left[-\frac 12 (\bq - \bQcal_n)\bGamma^{(n)}(\bq - \bQcal_n)\right],
\end{equation}
resulting in the following conditions that restrict the degrees of freedom of $\bQcal_n$ and $\bGamma^{(n)}$
\begin{equation}
    \bm S\bQcal_n = \bQcal_n, \qquad
    f_n(\bm q) = f_n(\bm S \bm q).
    \label{eq:symmetry:Q}
\end{equation}
The constraint on $\bQcal_n$ enforces the centroid of the network to be a high-symmetry point of the crystal (Wyckoff position), while the constraint on $f_n$ enforces the symmetry on the $\bGamma$ matrix as
\begin{equation}
    {\bm S}{}^\dagger \bm \Gamma_n \bm S = \bm \Gamma_n.
    \label{eq:symmetry:Gamma}
\end{equation}
These conditions coincide with the ones complied by $\bRcal$ and $\bPhi$ in the standard harmonic Hamiltonian to preserve the symmetry of the density matrix. Therefore, imposing symmetries on the Gaussian stretchers is equivalent to constraining symmetries of positions and dynamical matrices.

Imposing crystal symmetries by constraining each layer of the nonlinear transformation suppresses the size of the variational space spanned by the Gaussian stretchers. Therefore, we resort to this scheme only for symmetries that preserve the physical meaning of the resulting density matrix, like the exchange between indistinguishable particles and the total translational symmetry. Any further symmetry constraint is imposed \emph{a posteriori} throughout a Lagrange multiplier like
\begin{equation}
    \mathcal L = \sum_{i=1}^N\left\|\bm S \bm q^{(n)}(\bm q^{(1)}_i) - \bm q^{(n)}(\bm S \bm q^{(1)}_i)\right\|^2.
\end{equation}
By including $\mathcal L$ in the cost function to minimize the free energy, symmetries can be imposed a posteriori only between the first and the last layer of the \emph{Gaussia stretcher}, thus increasing the variational degrees of freedom of the solution.

Since multiple transformations can be stacked layer-by-layer, it is also convenient to employ a linear transformation. In normal circumstances, this is redundant with the SCHA itself. the SCHA centroids $\bRcal$ and auxiliary force constant matrix $\bPhi$ act as a linear transformation on an uncorrelated ensemble of normalized Gaussians. The transformation reads as
\begin{equation}
    \bm q^{(1)} = \bRcal + \bm L \bm q^{(0)} ,
\end{equation}
\begin{equation}
    \bm L \bm L^\dagger = \bUps^{-1},
    \label{eq:L:transf:def}
\end{equation}
where $\bUps^{-1}$ is the covariance matrix of the SCHA distribution
\begin{equation}
    \Upsilon_{ab} = \sqrt{m_am_b}\sum_\mu \frac{2 \omega_\mu}{2n_\mu + 1} e_\mu^a e_\mu^b.
\end{equation}
Using a first (or last) linear layer is a feature familiar to standard machine learning neural networks. 
Besides the resemblance with machine learning, using the linear transformation matrix $\bm L$ as a variational degree of freedom instead of the auxiliary force constant matrix $\bPhi$ ensures the continuity of the transformation of the ensemble even in the presence of mode crossing and degeneracies if the original ensemble $q^{(0)}$ is kept fixed. This is essential to have a smooth and differentiable energy landscape (cost function). 
Moreover, in the case of indistinguishable particles, this linear transformation can restore the harmonic correlation lost by considering $\bPhi^\text{(diff)}\to 0$ in \eqname~\eqref{eq:scha:electrons} to simplify the population of fermionic degrees of freedom.
Without loss of generality, $\bm L$ can be written as
\begin{equation}
    L_{ab} = \frac{1}{\sqrt m_a} \sum_\mu \sqrt{\frac{2\omega_\mu}{2n_\mu + 1}} e_\mu^a e_\mu^b,
\end{equation}
thus satisfying \eqname~\eqref{eq:L:transf:def}.


\section{Fermionic states}
\label{sec:electrons}

In this section, we present the constraints that the Gaussian stretcher have to satisfy to preserve the fermionic (or bosonic) character of the auxiliary density matrix $\tilde\rho(\bq, \bq')$ when it is transformed into the real space density matrix $\hat\rho(\bR, \bR')$.
This is the fundamental requirement to treat indistinguishable particles within the formalism presented in this work. In \secname~\ref{sec:harm:fermions}, we introduced the noninteracting fermionic wavefunction solving the auxiliary harmonic Hamiltonian. The manifold curvature mixes the degrees of freedom of different particles, leading to a correlated state. To preserve the antisymmetric characteristic of the underlying density matrix across the nonlinear transformation $\bR(\bq)$, the manifold should preserve the exchange operation between different particles. It is trivial to show that $\rho(\bR, \bR')$ is a antisymmetric if $\tilde\rho(\bq, \bq')$ is antisymmetric and $\bR(q)$ satisfies the condition
\begin{equation}
    {\bm R}_1(\bq_1, \bq_2, \cdots) = {\bm R}_2(\bq_2, \bq_1, \cdots),
    \label{eq:sym:transf}
\end{equation}
where with $\bq_1$ (and $\bR_1$) we indicate the components of $\bq$ (and $\bR$) on particle $1$, $1$ and $2$ are any two fermions, and the $\cdots$ indicates that the $\bq$ vectors of the other particles have not been exchanged.
\eqname~\eqref{eq:sym:transf} is fulfilled by a transformation of the kind:
\begin{equation}
    \bR(\bq) = \left\{
    \begin{array}{l}
    \displaystyle 
    \bR_1(\bq) = f(\bq) \bq_1\\
    \\
    \displaystyle
    \bR_2(\bq) = f(\bq) \bq_2\\
    \\
    \qquad\;\;\,\vdots
    \end{array}
    \right.
\end{equation}
where $f(\bq)$ is a generic symmetric function with respect to particle changes.
The Gaussian stretchers meet this condition if $\Qcal$ is the same for all electrons and most of $\bGamma$ parameters are constrained so that
\begin{equation}
    \Qcal_{(i,\alpha)} = \Qcal_{(j,\alpha)},
\end{equation}
\begin{equation}
    \Gamma_{(i,\alpha)(i,\beta)} = \Gamma_{(j,\alpha)(j,\beta)}, 
\end{equation}
\begin{equation}
    \Gamma_{(i,\alpha)(j,\beta)} = \Gamma_{(j,\alpha)(k,\beta)} ,
    \label{eq:gamma:cross}
\end{equation}
where the index $(i,\alpha)$ indicates atom $i$ and Cartesian coordinate $\alpha$,
for any choice of the atoms $i,j,k$. 
These are the same conditions as the $\bRcal$ and $\bPhi$ matrix of the auxiliary harmonic Hamiltonian for fermions we derived in \secname~\ref{sec:harm:fermions}. Thanks to the cross diagonal terms of $\bGamma$ (\eqname~\ref{eq:gamma:cross}), the transformation can still couple different particles, introducing correlations at each layer of the nonlinear transformation defining the curved manifold. 
Another consequence of the exchange symmetry of the manifold is the combinatorial number of constraints imposed on the degrees of freedom that counterbalance their growth when the number of particles increases: above two electrons, the degrees of freedom do not increase with the number of particles on each layer of the Gaussian stretcher.

\section{Applications and examples}
\label{sec:applications}
In this section, we illustrate some applications to show how the NLSCHA with the Gaussian stretcher systematically improves the SCHA result. We start with a one-particle problem, where deviations from the Gaussian ground state are significant, like in the case of a profound double-well potential for both the ground (\secname~\ref{sec:double:well}) and excited state (\secname~\ref{sec:excited}),  a Coulomb potential (hydrogen atom) in \secname~\ref{sec:hydrogen}, and finally for interacting electrons in the dissociation of the \ch{H2} molecule (\secname~\ref{sec:h2:dissociation}). We discuss and display also the capabilities of the method to capture excited states by exciting electrons in the auxiliary harmonic Hamiltonian for the 1D double potential and the \ch{H2} dissociation in the antibonding triplet electronic configuration.

\subsection{Double well potential}
\label{sec:double:well}
The double-well potential is a classic example of a strongly anharmonic system. It is also a good prototype for testing quantum tunneling, a regime in which the SCHA is known to fail\cite{siciliano_beyond_2024}.
We represent the double well potential of a 1D particle as
\begin{equation}
V(R) = a R^4 - bR^2.\label{eq:trial:pot}
\end{equation}
For values of $b > 0$, the harmonic approximation presents an imaginary frequency in $R = 0$, and the system becomes highly anharmonic.
Here, we compare the performance of the SCHA and the NLSCHA, with the multi-layer Gaussian stretcher manifold introduced in this work, against the exact (numerical) solution. 



\begin{figure*}
    \centering
    \includegraphics[width=\linewidth]{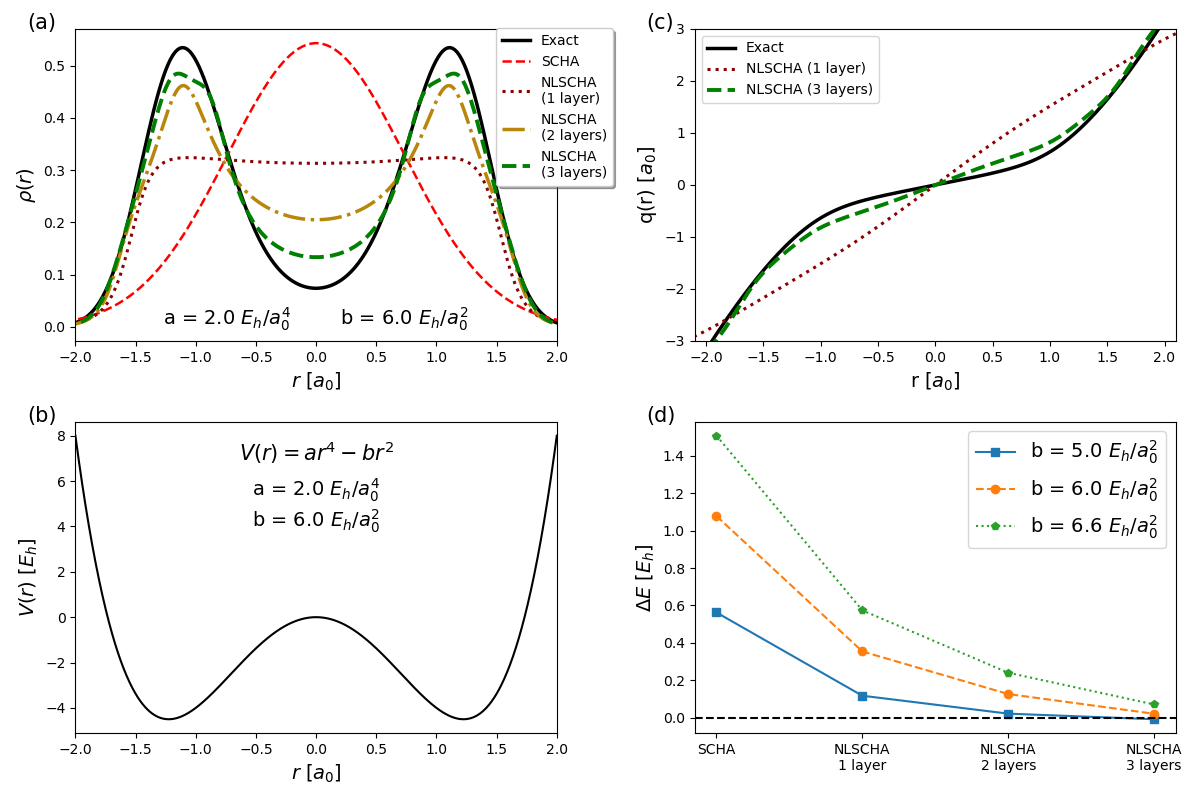}
    \caption{Comparison between the SCHA, NLSCHA, and exact (numerical) solution for a double well potential. (a) Probability density of the ground state wavefunction at $b = \SI{6.0}{\hartree\per\bohr^2}$ by varying the number of layers of the nonlinear transformation. (b) Potential energy landscape of the double well. (c) Comparison between the nonlinear transformations $q(r)$ encoded by the neural network and the one that exactly maps the SCHA Gaussian wavefunction to the true ground state of the double well potential.  
    (d) The energy error at several values of $b$ and varying the number of layers. All the values are in the deep quantum regime $b >\SI{2}{\hartree\per\bohr^2}$. The wavefunction is constrained to keep the inversion symmetry (inhibiting the transition into the localized state).
    The exact ground state of the double well potential has been evaluated numerically with the implicitly restarted Lanczos algorithm\cite{Lanczos} as implemented in Scipy\cite{scipy}.
    }
    \label{fig:double:well}
\end{figure*}

The double-well potential has three regimes as $b$ varies. When $b^3\gg\frac{2a^2}{m^2}$ (in Hartree atomic units), the two wells are far apart, the energy barrier is too high, and the solution localizes into one of the minima (broken symmetry solution). The SCHA describes this regime well if we allow the Gaussian solution to break the inversion symmetry and localize in one of the two minima, similarly to how unrestricted Hartree-Fock correctly captures molecular dissociation by breaking the spin symmetry. The opposite occurs when the barrier is small ($b^3\ll\frac{2a^2}{m^2}$); in this case, quantum/thermal fluctuations entirely overcome the energy barrier, and the probability density is located in the saddle point $R=0$. This phase is also well captured by the SCHA.
However, when the energy barrier is comparable with the quantum/thermal fluctuations ($b^3\approx\frac{2a^2}{m^2}$), tunneling and thermal hopping between the two minima occur on the same timescale as the lattice vibrations. This is the regime where we expect the SCHA to fail.
\figurename~\ref{fig:double:well}(a) reports the comparison between the ground state electronic densities of the SCHA and \name applying a different number of layers with the exact (numerical) solution. This test is performed for $b = \SI{6}{\hartree/\bohr^2}$, constraining the solution to satisfy the inversion symmetry (restricted SCHA). This potential is deep in the regime of quantum tunnelling between the two minima.

Applying multiple layers of the Gaussian stretcher improves the quality of the results. One layer of Gaussian stretcher is not enough to split the wavefunction into the two peaks around the minima of the potential, as it is energetically more convenient to reproduce the distribution's tails due to the fast rise of the quartic potential. However, adding extra layers enables the separation of the peaks. 
In this case, as with any 1D system, it is easy to show that the NLSCHA converges to the exact solution. In fact, we can prove that a nonlinear transformation $q(r)$ that maps a Gaussian into any possible ground state wavefunction always exists. 
In particular, $q(r)$ solves the differential equation
\begin{equation}
    \frac{dq}{dr} = \left(\frac{\psi(r)}{\tilde\phi[q(r)]}\right)^2,
    \label{eq:transform:diff}
\end{equation}
which is possible if $\psi(r)$ and $\phi[q(r)]$  share the same phase for all values of $r$ (apart from a global phase factor).
In 1D with the time-reversal symmetry, the ground state can always be chosen as a real positive function, and the square in \eqname~\eqref{eq:transform:diff} becomes the ratio between probability densities.
Then, the Bernier's theorem for optimal transport\cite{Brenier1991PolarFA} ensures that always exists a solution for \eqname~\eqref{eq:transform:diff}, and the $q(r)$ can be expressed as the gradient of a convex scalar function. In \figurename~\ref{fig:double:well}(c), we solved \eqname~\eqref{eq:transform:diff} for the map between the exact ground state of the double-well potential and the normal distribution, and compared to the neural network approximations of $q(r)$ obtained by variationally minimizing the ground state energy. 

In \figurename~\ref{fig:double:well}(d), we show the error on the ground state energy as a function of the number of layers and the $b$ parameter. We impose inversion symmetry, preventing the SCHA from localizing the Gaussian around one of the two minima, which is energetically more favorable for these values of $b$. 

The NLSCHA's ability to reproduce the splitting between wavefunction peaks and to minimize the error in the non-broken-symmetry solution has significant repercussions for the simulation of quantum tunneling, which was previously impossible within the standard SCHA or other harmonic-based approaches.

\subsection{Excited states}
\label{sec:excited}
\begin{figure}
    \centering
    \includegraphics[width=\linewidth]{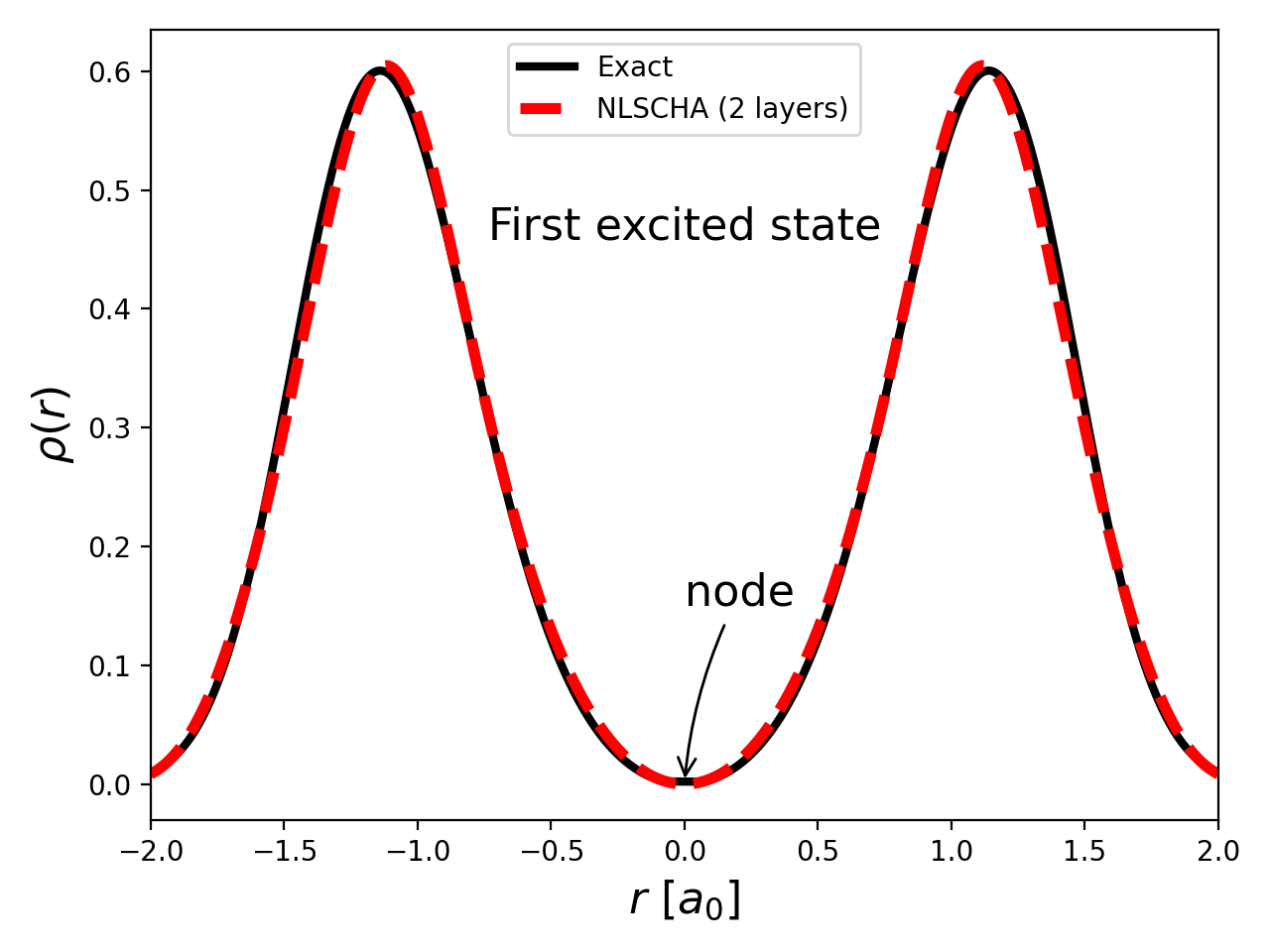}
    \caption{Comparison between the exact and \name (2 layers) excited state density. The potential is a double well of the form $V(r) = a r^4 - br^2$, where $a = \SI{2}{\hartree\per\bohr^4}$ and $b = \SI{-6}{\hartree\per\bohr^2}$, as in \figurename~\ref{fig:double:well}(c). We highlight the position of the node of the wavefunction in the origin.}
    \label{fig:excited:double:well}
\end{figure}

The new formalism introduced in this work also allows populating excited states in the auxiliary harmonic Hamiltonian, enabling the simulation of wavefunctions with nodes to optimize the total energy. This approach has already been successfully employed in variational Monte Carlo (VMC) to study optical gaps in solid state systems\cite {Cuzzocrea2020,Dash2019}, and transfers naturally to SCHA. As a simple benchmark, we apply the \name to unveil the wavefunction of the first excited state of the double well potential in the strongly anharmonic regime.
The comparison with the exact diagonalization is reported in \figurename~\ref{fig:excited:double:well}.
The wavefunction agrees exceptionally well with the exact result already with just two layers of Gaussian stretcher manifold, outmatching even the precision in the ground state thanks to the presence of a node in the original wavefunction.

\subsection{Hydrogen atom: the electron-ion cusp}
\label{sec:hydrogen}
While the \name performs exceptionally well on ionic systems and solves many issues of the SCHA, like molecular rotations, delocalization, and tunneling, its applicability in the electronic problem has to be proven. Electronic correlated wavefunctions are strongly non-Gaussian, with cusps in the overlap between different electrons and electron-ion due to the divergence of the Coulomb potential\cite{kato1957eigenfunctions}. While some of the problems could be accounted for with the use of pseudo-potentials, like done in mean-field approaches with a plane-wave basis set, in this section, we benchmark the expressibility of the transformed wavefunction on the hydrogen atom, a prototypical system with an electron-ion cusp that is highly challenging to all methods that do not explicitly account for the cusp in the wave-function (as it occurs in plane-wave codes). 

\begin{figure}
    \centering
    \includegraphics[width=\columnwidth]{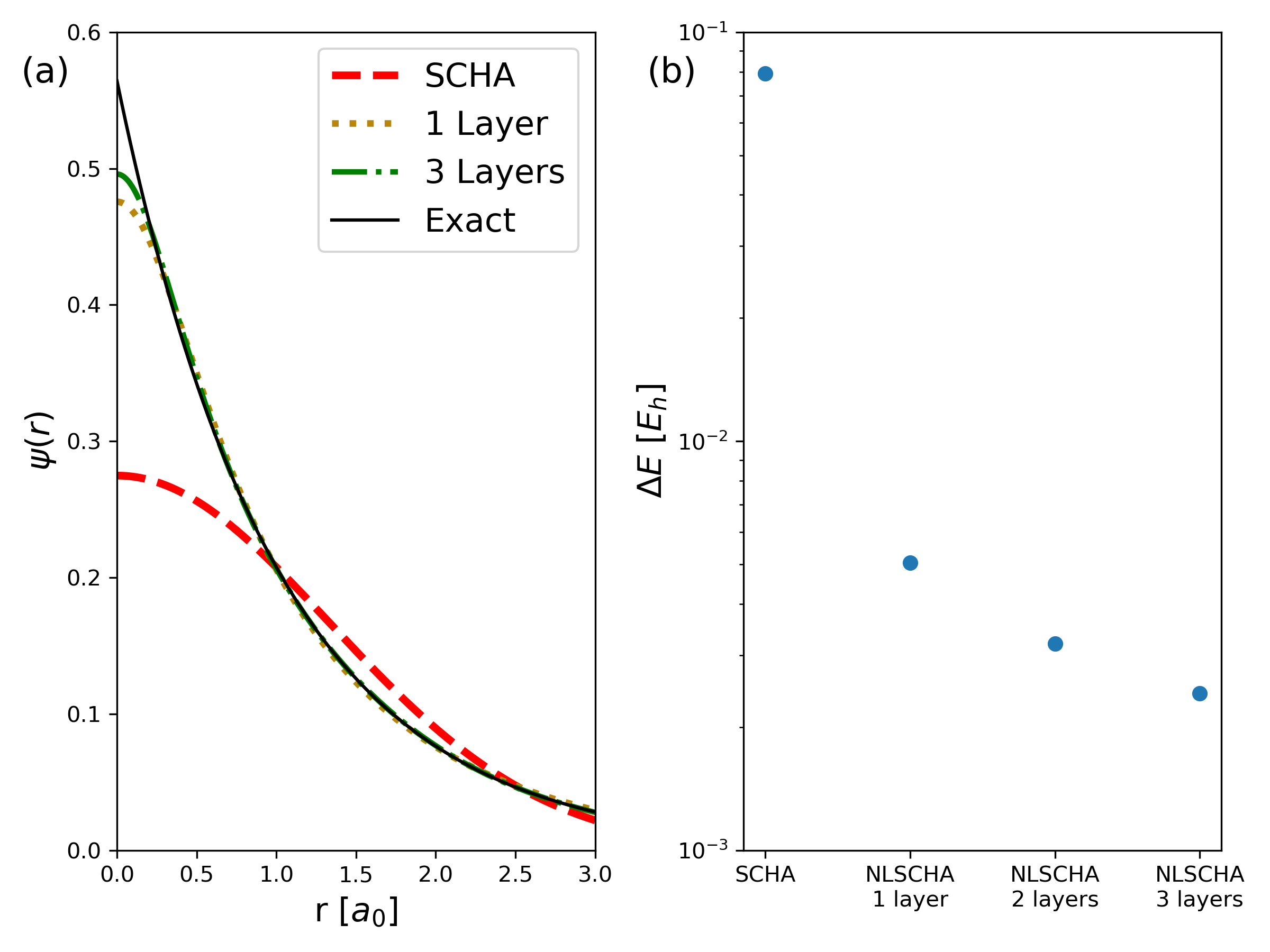}
    \caption{Hydrogen atom, comparison between the SCHA, nonlinear SCHA (\name), and exact (analytical) solution. (a) Comparison between the radial wavefunction of the $1s$ orbital around the electron-ion cusp in the origin. (b) Error on the ground state energy compared to the analytical result (log scale).}
    \label{fig:hydrogen:atom}
\end{figure}

We constrain the wavefunction for the hydrogen atom potential with a fully rotational symmetry group centered around the origin and the potential $V(r) = - 1/r$, as discussed in \secname~\ref{sec:neural:network}. We solve the system by progressively increasing the number of layers of the Gaussian stretchers. Thanks to the spherical symmetry constraints, each layer has only 2 degrees of freedom ($\zeta$ is free, $\Gamma$ is proportional to the identity matrix, and $\bQcal = 0$).
The final wave functions and energies are reported in \figurename~\ref{fig:hydrogen:atom}. Already with one layer, the long-range behavior of the wavefunction is very good ($r > \SI{0.5}{\bohr}$), while increasing the number of layers further improves the $r\to 0$ limit (the cusp).
In contrast to the 1D double-well potential—where adding additional layers leads to an exponential convergence toward the exact solution—we observe only more gradual improvement beyond the first transformation layer in \figurename~\ref{fig:hydrogen:atom} (b). This limited improvement rate is also evident in the corresponding wave-function profiles (\figurename~\ref{fig:hydrogen:atom}a), where the overall shape changes only slightly between one and three layers.
The underlying reason lies in the functional form of the Gaussian stretcher: its transformed wave function is inherently $C_\infty$, and thus a considerable number of layers would be required to reproduce the derivative discontinuity introduced by the divergent Coulomb potential. Nevertheless, the hydrogen atom's ground state can be expressed as a real wavefunction; thus, there exists a mathematical transformation that solves \eqname~\eqref{eq:transform:diff}, and the \name is exact in this case. However, achieving chemical accuracy requires overcoming this limitation of Gaussian stretchers, likely through \emph{ad hoc} nonlinear transformations that introduce controlled derivative discontinuities.

\subsection{\ch{H2} dissociation}
\label{sec:h2:dissociation}

We have tested the \name on problems with only one electron, where mean-field approaches are exact. Here, we tackle the \ch{H2} molecule under dissociation, a fully interacting many-body system, and one of the most challenging problems for mean-field approaches like Hartree-Fock, DFT, or many-body diagrammatic expansions\cite{Olsen2014,Giesbertz2018}. \ch{H2} is one of the few systems with strong electron-electron correlation that can be efficiently solved numerically, thus being the best benchmark for new methods. 
When the two H nuclei are far apart, the trivial solution (2 hydrogen atoms with electrons localized on different nuclei) is correlated, as one electron's location on an atom determines the atom on which the other electron must be. Thus, the ground state is composed of a linear combination of two Slater determinants.
Here, we benchmark the ground and the first excited state of the \ch{H2} during dissociation, preserving the correct spin state (singlet and triplet) along the process.


To model the \ch{H2} molecule, we employ a soft-core Coulomb potential that alleviates the electron-ion cusp and admits a nontrivial solution in one dimension while keeping all the essential correlation properties of the original Hamiltonian. 
The \ch{H2} soft-core Coulomb potential is composed of the one particle electron-ion interaction $V(r)$, the electron-electron interaction, and the ion-ion interaction (which is constant):
\begin{equation}
    V(r_1, r_2) = V(r_1) + V(r_2) + \frac{1}{\sqrt{(r_1 - r_2)^2 + \eta^2}} + \frac {1}{\sqrt{d^2 + \eta^2}}
    \label{eq:soft:coulomb}
\end{equation}
\begin{equation}
    V(r) = - \frac{1}{\sqrt{(r - d/2)^2 + \eta^2}} - \frac{1}{\sqrt{(r + d/2)^2 + \eta^2}} 
    \label{eq:soft:el}
\end{equation}
where $d$ is the inter-atomic distance, $\eta$ the softening factor of the Coulomb potential, and $r_1, r_2$ are the coordinates of the two electrons. It is known that the soft-Coulomb potential fundamentally changes the chemistry of bonding, as it affects the Coulomb potential even for $r \gg \eta$ \cite{SoftCoulomb2009,Iarrea2019}; however, in production simulations, this problem can be efficiently dealt with proper treatment of pseudo-potentials, and allows us to separate the numerical instabilities originating from the electron-ion cusp (treated explicitly in \secname~\ref{sec:hydrogen}) with the error of the method in describing electronic correlations.

\begin{figure}
    \centering
    \includegraphics[width=\columnwidth]{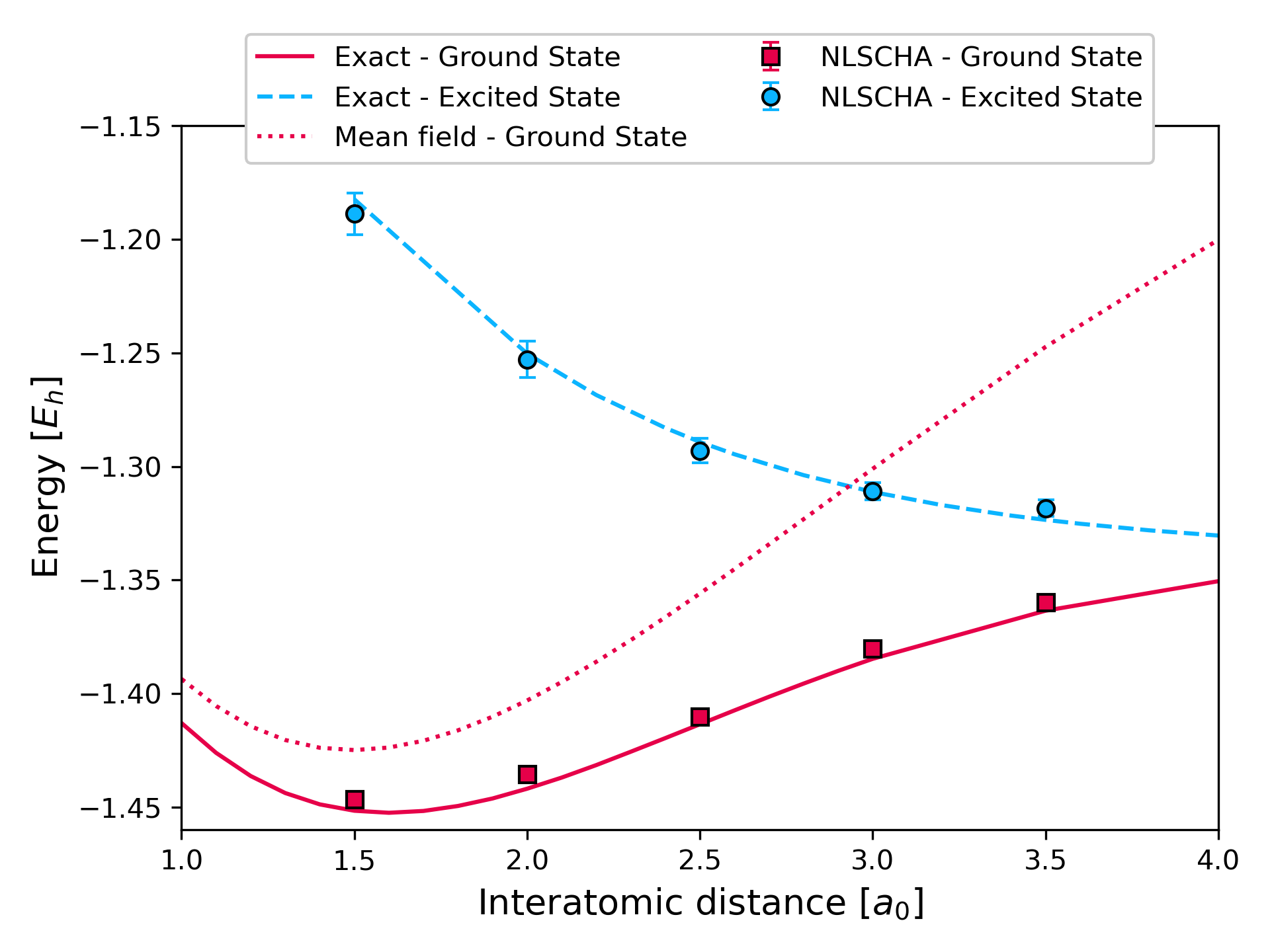}
    \caption{Dissociation energies of \ch{H2}. We compare the exact results obtained with numerical diagonalization of the Hamiltonian with the \name. The ground state is obtained in the singlet wave function, where both electrons have opposite spin in the ground state of the auxiliary harmonic oscillator. The excited state is the anti-bonding triplet solution, where one electron is in the ground state of the auxiliary harmonic Hamiltonian while the other is in the first excited state. The ground state's mean-field energy (Hartree-Fock) is also reported as a reference. The deviation in the dissociation limit to the expected energy for H$_2$ (-1.0 E$_h$) is due to the soft-core Coulomb potential (\eqname~\eqref{eq:soft:coulomb})
    \label{fig:h2:dissociation}}
\end{figure}

The results are reported in \figurename~\ref{fig:h2:dissociation}, where we compare in red the \name with a six-layer transformation to the numerical exact diagonalization and the mean-field Hartree-Fock (HF) solution.
As expected, HF reflects the problem of all mean-field approaches (as density functional theory) suffering from the static correlation error arising when the electrons localize on the atoms. The \name delivers an excellent agreement with the exact (numerical) result within the stochastic accuracy of the averages (\SI{0.001}{\hartree} in the ground state).

We benchmarked the antibonding excited state solution by promoting an electron to the first excited state of the auxiliary harmonic Hamiltonian in a triplet spin configuration. Notably, the \ch{H2} anti-bonding state is strongly affected by electron-hole interactions, as after the promotion of the electron in the excited state of the auxiliary Hamiltonian, the system significantly relaxes with dynamics similar to those giving rise to excitonic states in condensed matter systems. Therefore, the method displays promising potential for applications in materials with excitonic effects. Indeed, for a complete characterization of excitons, a dynamical extension of the method, in line with what is done for time-dependent SCHA\cite{Monacelli2021}, is required. 

\begin{figure*}
    \centering
    \includegraphics[width=0.8\textwidth]{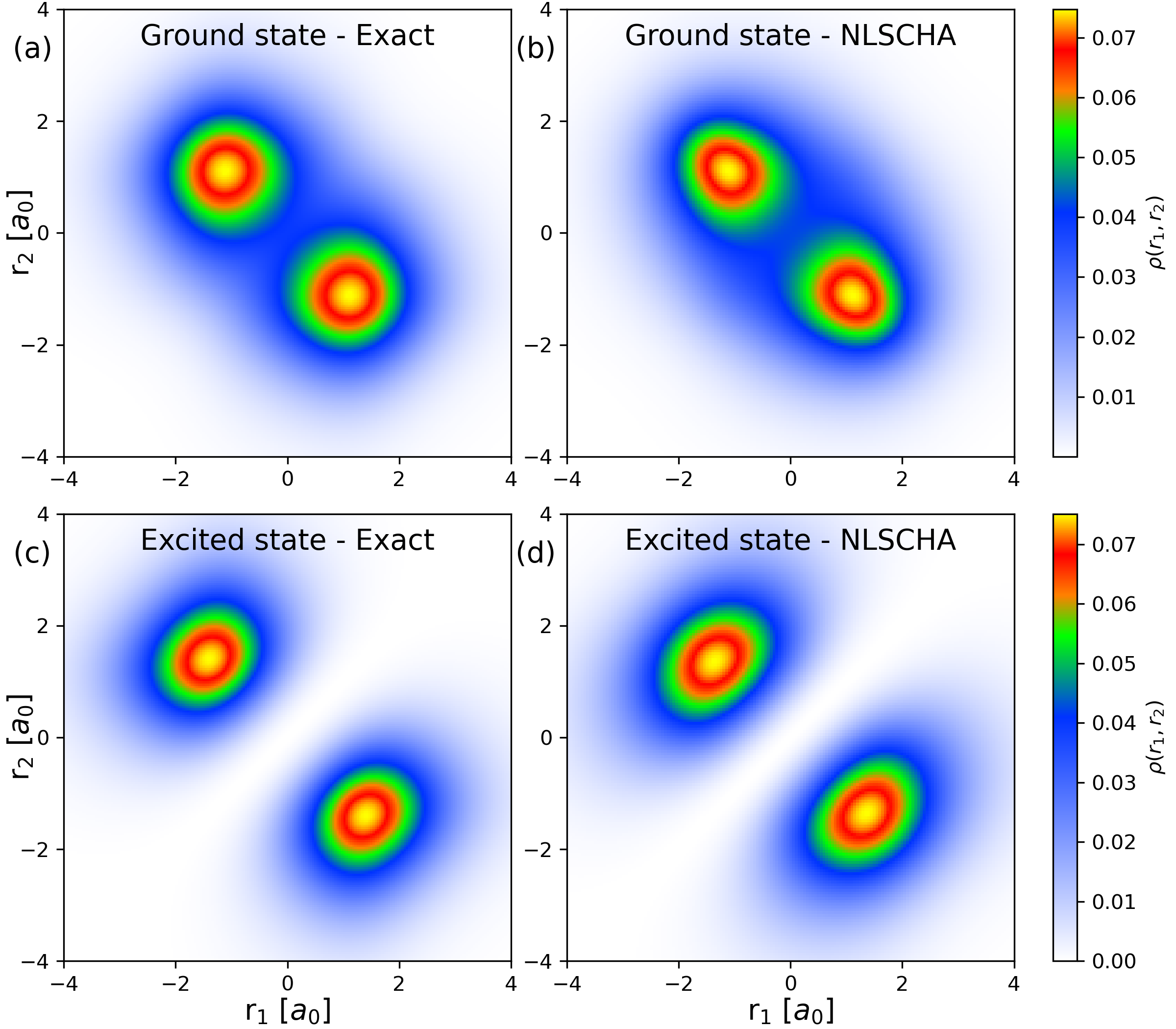}
    \caption{2-body electron density $\rho(r_1, r_2)$ for the \ch{H2} in dissociation ($d = \SI{2.5}{\bohr}$) in the 1D soft-core Coulomb potential. (a) The ground state density evaluated with exact (numerical) diagonalization. (b) Ground state density evaluated within the \name, employing 6 layers in the nonlinear transformation from the Gaussian. (c) First excited state with exact (numerical) diagonalization. This is an anti-bonding state with the spins in the triplet configuration. (d) Excited state density computed with the \name, employing 2 layers in the nonlinear transformation from the Gaussian.
    The error measured as the difference between the exact and NLSCHA probability density is reported in \appendixname~\ref{app:error}, \figurename~\ref{fig:h2:dissociation:wf:err}.
    \label{fig:h2:dissociation:wf}}
\end{figure*}

As the method is based on a first-principles quantization with a wavefunction approach, like VMC, we have an analytical expression for the resulting many-body wavefunction that we compare in \figurename~\ref{fig:h2:dissociation:wf} with the exact (full diagonalization) solution. The absence of spots in the electronic density located when $r_1=r_2$ over the hydrogen nuclei
confirms that the \name correctly accounts for static correlation, localizing the two electrons always in the two different ions. This is achieved thanks to the off-diagonal elements in the $\bGamma$ matrix of the Gaussian stretcher that couple the wavefunction of different electrons (\secname~\ref{sec:electrons}), like a backflow transformation employed in VMC.

\section{Limits of the current implementation}
\label{sec:limits}
In this section, we present the limits of the current implementation and discuss how they could be overcome.

As shown in \secname~\ref{sec:hydrogen}, the \name struggles to reproduce the electron-ion cusp, especially when large unscreened nuclear charges are in play (like for the case of core electrons), requiring a diverging number of hidden layers in the curved manifold. However, this is a common problem in electronic-structure simulations, e.g., in plane-wave basis sets, and has been addressed \emph{a posteriori} by employing pseudopotentials that smear out the electron-ion cusp near the nucleus.
Also, the electron-electron cusp is likely to lead to numerical instabilities, and it is less clear how to fix it. However, due to its repulsive nature, the contribution of the exact wavefunction shape impacts the total energy much less as the many-body electron density goes to zero on the cusp. An alternative approach would be to design an \emph{ad hoc} transformation whose Jacobian introduces a Jastrow-like factor that encodes the correct electron-ion and electron-electron cusp.

Different network architectures can be used to parameterize the curved manifold, like the Neural Ordinary Differential Equations (NODEs)\cite{NODEs} and eventually stacked on top of the Gaussian stretchers to improve accuracy.

Our current implementation does not yet support periodic boundary conditions. Periodic system of distinguishable particles (like atomic nuclei) can be trivially treated by enforcing translational equivariance in the Gaussian stretcher and periodicity of the auxiliary harmonic Hamiltonian by imposing \eqname~\eqref{eq:symmetry:Q} and \eqname~\eqref{eq:symmetry:Gamma}. These are precisely the same conditions on $\Gamma$ that a standard phononic force-constant matrix must satisfy in the presence of periodic boundary conditions.
However, these requirements become too strict when combined with the antisymmetry constraints discussed in \secname~\ref{sec:electrons}. 
To overcome this limitation for fermionic systems, the nonlinear transformation must be generalized, e.g., operating in relative coordinates. The proper definition and characterization of such a case go beyond the scope of this work.

Our framework allows, in principle, to describe systems at finite temperatures, as we optimize the density matrix to minimize the full Helmholtz free energy in the canonical ensemble. This is achieved thanks to the analytical expressions for the entropy in \eqname~\eqref{eq:S:phonons} and \eqref{eq:S:electrons}. However, while our mapping is exact for pure states (at least for real and positive wavefunctions), it is not at finite temperatures. In fact, the curved manifold maps uniformly all the excited states of the auxiliary Hamiltonian into the real Cartesian space without affecting their energy. Therefore, the energy spacing between excited states of the transformed auxiliary system is preserved, not matching in general the real Hamiltonian one. Since the entropy depends solely on the occupations of states, which rely on their relative energy, the excited states of the auxiliary and real Hamiltonian are populated differently.
To solve this problem, ref.\cite{siciliano_beyond_2024} proposed to introduce a new variational parameter: the mass-tensor of the auxiliary Hamiltonian. This mass tensor effectively operates a linear transformation on the momentum operator, providing a way to rescale the energy level of excited states in the auxiliary Hamiltonian to resemble that of the real Hamiltonian more closely. A different approach is to use all the occupation numbers $f_i$ in \eqname~\eqref{eq:S:electrons} as variational parameters, similarly to the scheme proposed for ions by ref.\cite{Zhang_solid_2025}. This introduces a much higher number of degrees of freedom (infinite, in principle, for every mode), whose optimization can be discontinuous in correspondence with avoided crossing and hybridization of excitations. However, despite the challenges of reaching an exact solution at finite temperatures, the nonlinear SCHA with a variational mass is already expected to yield excellent results, particularly for finite temperature electron-ion systems, as most thermal effects impact the ionic density matrix, for which even the standard SCHA provides a very good finite temperature description\cite{SSCHA}.

\section{Conclusions}
The introduction of the SCHA in a curved manifold enabled systematic improvement of the variational solution of the SCHA while preserving the analytical solution of the auxiliary Harmonic system. 
In this work, we established how to build a curved manifold that can be iteratively improved by adding multiple hidden layers, like in a deep neural network. 


We not only introduced a new method to simulate electrons and ions within the same theoretical framework, but we also unveiled its promising perspective to study correlated systems. The \name outperforms mean-field approaches by capturing static correlation without breaking the spin symmetry, both in the ground and the excited state of \ch{H2}.  The similarity of the nonlinear transformation to a backflow and its Jacobian to a Jastrow factor suggests that it can also efficiently describe dynamical correlations, as in established variational Monte Carlo approaches.
Ultimately, this work transforms a robust theory for atomic vibrations into a versatile framework for electronic structure, opening a new chapter in the unified, first-principles simulation of quantum matter.

\section*{Acknowledgments}
L. M. acknowledges the European Union and the H2020 program for funding this project under the MSCA-IF, project ID 101018714. This research was supported by the NCCR MARVEL, a National Centre of Competence in Research, funded by the Swiss National Science Foundation (grant number 205602).

\begin{appendices}

\section{Kinetic energy}
\label{app:kinetic}
Here, we derive the kinetic energy:
\begin{equation}
    K = - \sum_{i = 1}^N \int \prod_j  dR_j  \psi^\dagger (\bR) \frac{\partial^2 \psi}{\partial R_i^2}
\end{equation}
Integrating by parts, we get
\begin{equation}
    K = \sum_{i = 1}^N \int \prod_j  dR_j \frac{\partial \psi^\dagger}{\partial R_i} \frac{\partial \psi}{\partial R_i}
\end{equation}
We can bring the density outside as
\begin{equation}
K = \sum_{i = 1}^N \frac{1}{2m_i}\int \prod_j  dR_j \rho(R) \frac{\partial \ln \psi^\dagger}{\partial R_i}\frac{\partial \ln\psi}{\partial R_i}
\end{equation}
From which we have
\begin{equation}
    \ln\psi = \ln\tilde\psi + \frac 12 \ln \det\bJ
\end{equation}
\begin{align}
\frac{d\ln\psi}{dR_a} &= \sum_b  \frac{d\ln\tilde \psi}{dq_b} \frac{dq_b}{dR_a} + \frac12\frac{d\ln\det\bJ}{dR_a} = \nonumber \\ 
&=\sum_b  \frac{d\ln\tilde \psi}{dq_b}J_{ba} + \frac 12 \sum_{cd}\Jcal_{cd}J^d_{ca}
    \label{eq:dlnpsi}
\end{align}
The overall kinetic energy is:
\begin{align}
    K = \sum_{i = 1}^N \frac{1}{2m_i}&\bigg[
    \sum_{bc}\left<J_{bi}J_{ci}\frac{d\ln{\tilde \psi}^\dagger}{dq_b}\frac{d\ln\tilde \psi}{dq_c}\right>  \nonumber \\
    &+ \frac 12 \left<\left(\sum_{cd}\Jcal_{cd}J^d_{ci}\right)\sum_e\frac{d\ln\tilde \rho}{dq_e}J_{ei}\right> \nonumber \\ 
    & + \frac 14\left<\left(\sum_{cd}\Jcal_{cd}J^d_{ci}\right)^2\right>
    \bigg],
    \label{eq:K:total}
\end{align}
where $\tilde\rho(\bm q) = \tilde\psi^\dagger(\bm q)\tilde\psi(\bm q)$. From \eqname~\eqref{eq:K:total}, we get the three observables:
\begin{equation}
    \alpha_i(\bq) = \frac 14 \left[\sum_{cd}\Jcal_{cd}(\bq)J^d_{ci}(\bq)\right]^2
\end{equation}
\begin{equation}
    \beta_i^e(\bq) =  J_{ei}(\bq)\left[\sum_{cd}\Jcal_{cd}(\bq)J^d_{ci}(\bq)\right]
\end{equation}
\begin{equation}
    \gamma^{bc}_i(\bq) = J_{bi}(\bq)J_{ci}(\bq)
\end{equation}

by exploiting the relation \eqname~\eqref{eq:jcal2}, we get:
\begin{equation}
    \sum_{cd} \Jcal_{cd}J^d_{ci} = -
    \sum_{ahk} J_{ha} J_{ki} \Jcal^a_{kh}
\end{equation}

This, indeed, works only for T = \SI{0}{\kelvin}; however, a similar approach can be derived also at finite temperature recognizing that:

\begin{equation}
    K = -\sum_i\frac{1}{2m_i}\int \prod_jdR_j dR_j' \delta(\bR - \bR') \frac{\partial^2 \rho(\bR, \bR')}{\partial {R'}_i^2}
\end{equation}
We can then exploit the integration by parts so that
\begin{equation}
    K = \sum_i\frac{1}{2m_i}\int \prod_jdR_j dR_j' \frac{\partial \delta(\bR - \bR')}{\partial R'_i} \frac{\partial \rho(\bR, \bR')}{\partial R'_i}
\end{equation}
Now we can exchange the derivative in the Dirac $\delta$ with $R$:
\begin{equation}
     K = -\sum_i\frac{1}{2m_i}\int \prod_jdR_j dR_j' \frac{\partial \delta(\bR - \bR')}{\partial R_i} \frac{\partial \rho(\bR, \bR')}{\partial R'_i}
\end{equation}
And integrating back by parts
\begin{equation}
     K = \sum_i\frac{1}{2m_i}\int \prod_jdR_j dR_j' \delta(\bR - \bR') \frac{\partial^2 \rho(\bR, \bR')}{\partial R'_i\partial R_i}.
\end{equation}

This expression allows us to recover a symmetry and a single derivative on each variable that we can exploit:
\begin{align}
    \frac{\partial^2 \rho(\bR, \bR')}{\partial R'_i\partial R_i} &= \sum_h \lambda_h \frac{\partial \psi_h}{\partial R_i}(\bR)\frac{\partial \psi_h^\dagger}{\partial R_i}(\bR') \nonumber \\
    &= \sum_h \lambda_h \left|\psi_h\right|^2\frac{\partial \ln\psi_h}{\partial R_i}(\bR)\frac{\partial \ln\psi_h^\dagger}{\partial R_i}(\bR')
\end{align}

Now, the same procedure as in \eqname~\eqref{eq:dlnpsi} can be exploited, where we end up with the same integrals as in \eqname~\eqref{eq:K:total}, but where the averages are taken over each state in the mixture defined by the density matrix.
\begin{align}
    K &= \sum_{i = 1}^N \frac{1}{2m_i}\bigg[
    \sum_{bc}\int d\bq J_{bi}J_{ci}\sum_h\lambda_h|\tilde\psi_h|^2\frac{d\ln{\tilde \psi_h}^\dagger}{dq_b}\frac{d\ln\tilde \psi_h}{dq_c}  \nonumber \\
    &+ \int d\bq\frac 12\left(\sum_{cd}\Jcal_{cd}J^d_{ci}\right)\sum_eJ_{ei}\sum_h\lambda_h|\tilde\psi_h|^2\frac{d\ln\tilde\rho_h}{dq_e} \nonumber \\ 
    & + \frac 14\int d\bq \sum_h\lambda_h|\tilde\psi_h|^2\left(\sum_{cd}\Jcal_{cd}J^d_{ci}\right)^2
    \bigg],
    \label{eq:K:total}
\end{align}
where
\begin{equation}\sum_h\lambda_h|\tilde\psi_h|^2\frac{d\ln{\tilde \psi_h}^\dagger}{dq_b}\frac{d\ln\tilde \psi_h}{dq_c} = \int d\bq'\delta(\bq - \bq')\frac{\partial^2 \tilde\rho(\bq, \bq')}{\partial q_b\partial q_c'}
\end{equation}
\begin{equation}
\sum_h\lambda_h|\tilde\psi_h|^2\frac{d\ln\tilde \rho_h}{dq_e} = \tilde\rho(\bq)\frac{d\ln\tilde\rho}{dq_e}
\end{equation}
\begin{equation}
    \sum_h\lambda_h|\tilde\psi_h|^2 = \tilde\rho(\bq)
\end{equation}
\begin{align}
    K = \sum_{i = 1}^N \frac{1}{2m_i}&\bigg[
    \sum_{bc}\int d\bq J_{bi}J_{ci}\int d\bq'\delta(\bq - \bq')\frac{\partial^2 \tilde\rho(\bq, \bq')}{\partial q_b\partial q_c'} \nonumber \\
    &+ \frac 12\int d\bq\left(\sum_{cd}\Jcal_{cd}J^d_{ci}\right)\sum_eJ_{ei}\tilde\rho(\bm q)\frac{d\ln\tilde\rho}{dq_e} \nonumber \\ 
    & + \frac 14\left<\left(\sum_{cd}\Jcal_{cd}J^d_{ci}\right)^2\right>
    \bigg],
    \label{eq:K:total}
\end{align}
\eqname~\eqref{eq:K:total} is the complete expression for the free energy, regardless of the auxiliary density matrix of choice $\tilde\rho(\bq, \bq')$. 
In the particular case of the Gaussian density matrix
\begin{align}
    \int d\bq'\delta(\bq - \bq')&\frac{\partial^2 \tilde\rho(\bq, \bq')}{\partial q_a \partial q_b} = \tilde\rho(\bq) \bigg(A_{ab} + \nonumber \\
    & + \sum_{cd} \frac{\Upsilon_{ac}\Upsilon_{bd}}{4} (q_c - \mathcal Q_c)(q_d - \mathcal Q_d)\bigg),
\end{align}
\begin{equation}
     \rho(\bq)\frac{d\ln \tilde\rho}{dq_e} = -\tilde\rho(\bq)\sum_b \Upsilon_{eb}(q_b - \mathcal Q_b),
\end{equation}
that can be fit into the expression of \eqname~\eqref{eq:K:total} to evaluate numerically the kinetic energy.

\section{High-order jacobians of the single layer Gaussian Stretcher}
To compute explicitly all the expression for the kinetic energy and the free energy gradients in the specific case of the Gaussian Stretcher transformation, we need to compute the Jacobian matrix up to the fourth order.
We start with the single Gaussian stretcher
\begin{equation}
    \Jcal_{ab} = \frac {dR_a}{dq_b} 
\end{equation}
\begin{align}
    \Jcal_{ab} &= \delta_{ab}\left[1 + (e^\zeta - 1)e^{ - \frac 12\tilde {\bq} \bGamma\tilde \bq }\right] \nonumber \\
    & - (e^\zeta - 1) e^{-\frac 12\tilde {\bq} \bGamma\tilde \bq}\sum_c\Gamma_{bc} (q_c - \Qcal_c)(q_a - \Qcal_a) 
    \label{eq:J}
\end{align}
We can go on with further derivatives.
\begin{equation}
    \Jcal^a_{bc} = \frac{dR_a}{dq_bdq_c}
\end{equation}
\begin{align}
\frac{\Jcal^a_{bc}e^{\frac 12\tilde {\bq} \bGamma\tilde \bq}}{ (e^\zeta - 1) } &= -\delta_{ab}\sum_h\Gamma_{ch} (q_h - \Qcal_h)
 -\delta_{ac}\sum_h\Gamma_{bh} (q_h - \Qcal_h)\nonumber \\ 
& + \sum_h\Gamma_{bh} (q_h - \Qcal_h)\sum_k\Gamma_{ck} (q_k - \Qcal_k)(q_a - \Qcal_a) \nonumber \\
& - \Gamma_{bc}(q_a - \Qcal_a)
\label{eq:J2}
\end{align}
\begin{align}
\frac{\Jcal^a_{bcd}e^{\frac 12\tilde {\bR} \bGamma\tilde \bR}}{ (e^\zeta - 1) } &= -\delta_{ab}\Gamma_{cd} - \delta_{ac}\Gamma_{bd}
\nonumber \\
&-\delta_{ad}\Gamma_{bc} + \delta_{ad}(\bGamma\tilde\bq)_b(\bGamma\tilde\bq)_c \nonumber \\
&+\tilde q_a \Gamma_{bd}(\bGamma\tilde\bq)_c + 
\tilde q_a \Gamma_{cd}(\bGamma\tilde\bq)_{b}
  \nonumber \\
& + \delta_{ab}(\bGamma\tilde\bq)_d(\bGamma\tilde\bq)_c+ \delta_{ac}(\bGamma\tilde\bq)_b(\bGamma\tilde\bq)_d \nonumber \\
& +\tilde q_a \Gamma_{bc}(\bGamma\tilde\bq)_d
- \tilde q_a (\bGamma\tilde\bq)_b(\bGamma\tilde\bq)_c(\bGamma\tilde\bq)_d
\end{align}

\section{high-order jacobians of the multilayer Gaussian stretcher}
\label{app:gauss:combined}
As suggested in the main text, nonlinear transformations can be concatenated to form more complex transformations. We can derive the jacobian by using the composite transformation:
\begin{equation}
    \frac{d}{dq_j}q^{(m)}_i(q_1^{(m-1)}, \cdots, q_n^{(m-1)}) = \sum_{a = 1}^n \frac{dq^{(m)}_i}{dq_a^{(m-1)}} \frac{dq_a^{(m-1)}}{dq_j}
\end{equation}
From which we get a recursive equation 
\begin{equation}
    \Jcal_{ij}^{(m)} = \sum_a  \Jcal_{ia}(\bq^{(m-1)}) \Jcal_{aj}^{(m-1)}(\bq^{(1)})
\end{equation}
Which means that the jacobian is the matrix product of the jacobians after each iteration.
The second order jacobian becomes
\begin{align}
    (\Jcal^{(m)})^i_{jk} =&\sum_{ab} \Jcal^i_{ab}(\bq^{(m-1)})\Jcal^{(m-1)}_{aj}\Jcal^{(m-1)}_{bk} +\nonumber \\
    &\sum_a \Jcal_{ia}(\bq^{(m-1)}) (\Jcal^{(m-1)})^a_{jk}
\end{align}

And the higher-order derivative becomes
\begin{align}
    (\Jcal^{(m)})^i_{jkh} =& 
    \sum_{abc} \Jcal^i_{abc}(\bq^{(m-1)}) \Jcal^{(m-1)}_{aj}\Jcal^{(m-1)}_{bk}\Jcal^{(m-1)}_{ch} + \nonumber \\
    & \sum_{ab} \Jcal^i_{ab}(\bq^{(m-1)})\left(\Jcal^{(m-1)}\right)^a_{jh}\Jcal^{(m-1)}_{bk} +\nonumber \\
    & \sum_{ab}\Jcal^i_{ab}(\bq^{(m-1)})\Jcal^{(m-1)}_{aj}\left(\Jcal^{(m-1)}\right)^b_{kh} +\nonumber \\
    & \sum_{ab}\Jcal^i_{ab}(\bq^{(m-1)})\Jcal^{(m-1)}_{bh} (\Jcal^{(m-1)})^a_{jk} + \nonumber \\
    &\sum_a \Jcal_{ia}(\bq^{(m-1)}) \left(\Jcal^{(m-1)}\right)^a_{jkh}
\end{align}

This chain rule allows the computation of all the jacobians as a forward propagation algorithm. This scheme can be seen as a neural network with a gaussian activation function.

\section{Some useful relations}

We simplify some explicit multiplication between Jacobians that often appear in the derivations.
For example, in the expression for $\beta^i_a(\bq)$ we find the multiplication:
\begin{equation}
\sum_{cd}\Jcal_{cd}J^d_{ca} = - \sum_{hkl} J_{kh}J_{la}\Jcal^h_{kl}
\label{eq:jcal2}
\end{equation}
In the same way, for $\alpha$ is important the following multiplication
\begin{align}
    \sum_{bc}\Jcal_{bc}J^c_{baa} = -\sum_{hkl}&\bigg[
    J_{kh}J_{la}\sum_m \Jcal^h_{lkm}J_{ma} + \nonumber \\
    & +2J^k_{ha}J_{la}\Jcal^h_{lk} + 
    J_{kh} J^l_{aa}\Jcal^h_{lk}
\bigg]
\end{align}

\section{Scaling}

The computational scaling of the free energy optimization algorithm with respect to the number of degrees of freedom, n, is a critical aspect of its performance. The dominant cost arises from the kinetic energy term, which depends on the Jacobian derivatives of the multi-layer coordinate transformation.
While these terms are in principle highly costly to compute and store in memory, they can be systematically decomposed into a series of multiplications between $n\times n$ matrices, where $n$ is the number of degrees of freedom, as is evident from the expressions in Eq.~\eqref{eq:J} and Eq.~\eqref{eq:J2}. The scaling is therefore dictated by the complexity of these underlying matrix operations.
For a general Gaussian stretcher where the covariance matrix $\Gamma$ is assumed to be dense, the resulting Jacobian and its derivative-related matrices are also dense. Consequently, fundamental operations such as matrix inversion or multiplication have a computational cost of $O(n^3)$. As shown by substituting Eq.~\eqref{eq:J2} into Eq.~\eqref{eq:jcal2}, even the inverse of higher-order tensors can be constructed from these $n\times n$ matrix products.
A multi-layer architecture composed of $m$ chained transformations requires these matrix operations to be performed for each layer (see Appendix~\ref{app:gauss:combined}), introducing a linear scaling with m. The optimization step, which relies on backpropagation to compute gradients, shares the same asymptotic complexity as the forward pass. Therefore, the overall theoretical cost of the algorithm scales as:
$$
O(m\cdot n^3)
$$
Notably, this scaling is comparable to the best \emph{ab initio} techniques, such as DFT and VMC.  
This scaling analysis hinges on the critical assumption of a dense $\Gamma$ matrix. It is plausible that employing a sparse or structured representation for $\Gamma$ could significantly reduce the computational cost, potentially leading to a more favorable scaling law.

\section{Anharmonicity in the double-well potential}

The anharmonicity in the double-well potential (\eqname~\ref{eq:trial:pot}) is introduced by the presence of the quartic term $R^4$, which has $a$ as a coefficient.
However, the lowest order correction to the average curvature of the potential is
\begin{equation}
    \left<\frac{d^2V}{dR^2}\right> = 12a\sigma^2 - 2b,
\end{equation}
where $\sigma^2 = \braket{RR}$.
Within the Harmonic approximation, we have at $T=\SI{0}{\kelvin}$
$$
\sigma^2 = \frac{\hbar}{2m\omega},\qquad \omega^2 = -\frac{b}{m}.
$$
This result indicates that the lowest-order correction to the harmonic frequency due to anharmonicity diverges as $b \to 0$, signaling the onset of a nonperturbative regime for positive values of $b$. For this reason, we analyze the double-well potential as a function of $b$ rather than $a$: the parameter $b$ more directly quantifies the degree of anharmonicity and represents a more demanding case for theoretical analysis.

\section{Error on the \ch{H2} density}
\label{app:error}
Here we report the error for the calculation of the \ch{H2} density.
The error is evaluated as 
\begin{equation}
    Err(r_1, r_2) = \left| \left|\Psi_\text{exact}(r_1, r_2)\right|^2 - \left|\Psi_\text{NLSCHA}(r_1, r_2)\right|^2\right|
    \label{eq:error}
\end{equation}

\begin{figure}
    \centering
    \includegraphics[width=\linewidth]{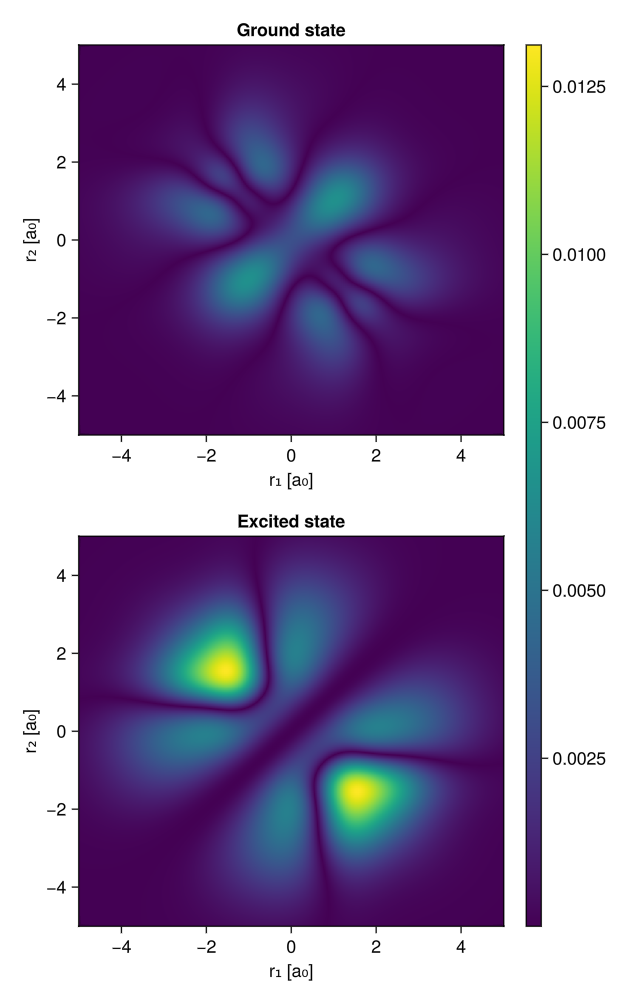}
    \caption{Error on the probability density of the NLSCHA wavefunction compared to the exact result for \ch{H2} with interatomic distance of \SI{2.5}{\bohr} (\figurename~\ref{fig:h2:dissociation:wf}). The error is evaluated using \eqname~\eqref{eq:error}.}
    \label{fig:h2:dissociation:wf:err}
\end{figure}

The results for the ground and excited states corresponding to the density reported in \figurename~\ref{fig:h2:dissociation:wf} are presented in \figurename~\ref{fig:h2:dissociation:wf:err}. 

\section{Harmonic oscillator of Fermions}
\label{app:harmonic:fermions}
Here, we report the solution of the harmonic Hamiltonian for fermions. The system is composed of $N$ fermions interacting within a harmonic potential. For simplicity, we consider a one-dimensional space. Due to the exchange symmetry, in 1D there are only 2 degrees of freedom: the self-interaction $a$ and electronic repulsion $b$
\begin{equation}
    V(q_1, \cdots, q_N) = 
    \frac 12 a \sum_{i = 1}^N
    q_i^2 - \frac 12 b
    \sum_{ij}(q_i - q_j)^2.
\end{equation}
Reorganizing the terms of the equation, we get
\begin{equation}
    V(q_1, \cdots, q_N) = 
   \frac 12  \left(a - b\right) \sum_{i = 1}^N
    q_i^2 - b
    \sum_{i\neq j}q_iq_j,
\end{equation}
\begin{equation}
    V(\bq) = \frac 12 \bpm 
    q_1 & \cdots & q_N\epm 
    \bpm (a-b) & -b & \cdots & -b \\
    -b & (a-b) & \cdots & -b \\
    \vdots & \vdots& \ddots & \vdots \\
    -b & \cdots & -b & (a-b) \epm 
    \bpm q_1 \\ \vdots \\ q_N \epm
\end{equation}
This is an interacting problem that can be trivially solved by diagonalizing the matrix. The solution have 2 eigenvalues
\begin{equation}
    \lambda_1 = a - Nb \qquad 
    \boldsymbol{e_1} = \bpm 1 \\ \vdots \\ 1\epm ,\label{eq:sym}
\end{equation}
\begin{equation}
    \lambda_2,\cdots,\lambda_n = a \qquad 
    \boldsymbol{e_2} = \bpm 1 \\ -1 \\ 0 \\  \vdots \\ 0\epm , \cdots,\boldsymbol{e_N} = \bpm 1 \\ 0\\  \vdots \\ 0 \\ -1\epm.
    \label{eq:exchange}
\end{equation}
The eigenvector $\boldsymbol{e_1}$ is already symmetric under the exchange of any couple of particles, each $\boldsymbol{e_i}$ with $i\ge 2$ is antisymmetric for exchanging two specific particles and symmetric for exchanging any other 2. Unfortunately, for more than 2 particles, it is impossible to build a basis from $\boldsymbol{e_2},\cdots,\boldsymbol{e_n}$ that commutes with all the exchange operators $\boldsymbol{S_{ij}}$. While we know the spectrum of the problem, this makes it very difficult to find exact selection rules to populate states in the case of fermions and bosons for more than 2 particles, in particular, if we also account for the spin degrees of freedom.
Another problem arises in the thermodynamic limit $N\rightarrow \infty$. In this case, the spectrum, and in particular $\lambda_1$ (\eqname~\ref{eq:sym}), becomes not positive definite for any small positive value of $b$. Thus, even if the Harmonic Hamiltonian may describe the correlation of molecules or systems with a finite number of electrons, $b\rightarrow 0$ in the thermodynamic limit, we recover a noninteracting Hamiltonian. For this reason, in the main text, we restrict to the noninteracting case where $b=0$.

\section{Position dependent effective mass tensor}
\label{app:mass}
As in classical Lagrange equations, the constrained motion on a curved manifold affects the effective masses of the systems. In classical mechanics, the kinetic energy after the change of variable is obtained as:

\begin{equation}
    K_\text{cl} = \sum_i\frac{1}{2m_i}\left< J_{ai} J_{bi} p_a p_b\right> 
    \label{eq:k:cl:primitive}
\end{equation}
where $p_a$ is the $a$-th component of the moment of the auxiliary variable $q_a$. The transformation of the variable thus can be seen as a mass that depends on the metric tensor:
\begin{equation}
    M^{-1}_{ab} = \sum_i\frac{J_{ai} J_{bi}}{m_i}
\end{equation}
\begin{equation}
    K_\text{cl} = \frac 12\sum_{ab} \left< p_a M^{-1}_{ab} p_b\right>.
    \label{eq:k:cl}
\end{equation}

Unfortunately, in quantum mechanics, it is not possible to quantize directly the auxiliary variables and thus express everything as a simple change of masses. However, we can always write the kinetic energy as 
\begin{equation}
    K = K_\text{cl} + K_\text{cor}
\end{equation}
where $K_\text{cor}$ is the quantum correction to the kinetic energy. In particular, from \eqname~\eqref{eq:kin}, the term that contains \eqname~\eqref{eq:k:cl} is the one that multiplies the $\bm \gamma$ tensor. 
\begin{align}
K_\text{cor} =  -\sum_{a = 1}^{3N}&\frac{\hbar^2}{2m_a}
    \left<\alpha^a(\bq) + \sum_i \beta^a_i(\bq)\hat P_i\right>
    \label{eq:k:cor}
\end{align}

Since in practice, the classical kinetic energy is both the biggest part of the kinetic energy and the one with the highest noise, it is convenient to rewrite it as the masses are constant (linear transformation) and average only the curvature:
\begin{equation}
    \bar M^{-1}_{ab} = \left<M^{-1}_{ab}(\bm q)\right>\qquad 
    \Delta M^{-1}_{ab}(\bm q) = M^{-1}_{ab}(\bm q) - \bar M^{-1}_{ab}
\end{equation}
\begin{equation}
    K_\text{cl} = \frac 12 \sum_{ab}\bar M^{-1}_{ab}
    \left<p_a p_b\right> + \frac 12\sum_{ab} \left< p_a\Delta M^{-1}_{ab}(\bm q) p_b\right>.
    \label{eq:k:cl:final}
\end{equation}
where the first term is analytical when for Gaussian wavefunctions in the auxiliary system.
The evaluation of the classical kinetic energy through \eqname~\eqref{eq:k:cl:final} strongly suppresses the stochastic noise compared with \eqname~\eqref{eq:k:cl:primitive}, but it has the same expected value.


\section{Fermionic and bosonic wavefunction}

Up to two electrons, the wave function remains symmetric as it can be described by a singlet state, where the antisymmetric part of the wave function is trivially encoded by their opposed spin. However, the factorization of space and spin for the exchange symmetry is only possible for two electrons. 
Therefore, to solve a realistic system beyond the Helium atom or the \ch{H2} molecule it is necessary to devise a new strategy to encode the antisymmetry of the wavefunction.

In the opening of \secname~\ref{sec:electrons}, we derived the constraint on the nonlinear transformation to keep the exchange symmetry of the wavefunction in the auxiliary variables throughout the nonlinear transformation operated by the neural network. Here, we devise a way to account for a fermionic (or bosonic) wavefunction directly in the auxiliary space.

The only requirements is to be able to:
\begin{itemize}
    \item Extract random configurations according to the correct statistics.
    \item Evaluate the modified kinetic energy average in the auxiliary space (\eqname~\ref{eq:k:cor} and \ref{eq:k:cl})
\end{itemize}

To extract random configurations distributed according to the Fermi-Dirac, we employ a standard quantum Monte Carlo algorithm sampling the wavefunction $\Psi(q)$ as the Slater determinant of the harmonic eigenfunctions.

Storing for each configuration in $q$ space the gradient of the wavefunction, we can easily compute the averages of kinetic operators that depends on both $P$ and $q$:

\begin{equation}
    i\hbar \left<\beta_a(\bm q) P_a\right> = -\hbar^2\left<\beta_a(\bm q) \frac{\partial\ln\Psi}{\partial q_a} \right>
    \label{eq:bPf}
\end{equation}
\begin{equation}
    \left<P_a\gamma_{ab}(q) P_b\right> = \hbar^2\left<\gamma_{ab}(\bm q)\frac{\partial\ln\Psi^*}{\partial q_a}\frac{\partial\ln\Psi}{\partial q_b} \right>
\end{equation}
Which is equivalent to substituting the ${\bm \Upsilon} \bq$ with $\nabla_{\bq}\ln\Psi$ from the equations of the standard SSCHA.
Interestingly, the derivative of the logarithm of the wavefunction is independent on its global phase, thus, \eqname~\eqref{eq:bPf} satisfies the Gauge invariance. Indeed, if the global phase depends on the position, care must be taken as \eqname~\eqref{eq:bPf} could be related to topological phenomena. Therefore, we must store, for each point in the auxiliary ensemble, the value of the gradient of the logarithm of the slater determinant, which can be evaluated efficiently using algorithmic differentiation. The analytical value of the kinetic energy in a pure state can be evaluated as
\begin{equation}
    \braket{P_aP_b} = \frac{1}{N_e} \delta_{ab}\frac{\hbar}{2} \sum_{\mu=0}^\infty \left[2n_\mu + 1\right]f_\mu
\end{equation}
where $f_\mu$ now indicate the number of electrons occupying the $n_\mu$ state of the auxiliary harmonic Hamiltonian along the $a$ or $b$ direction. 

\end{appendices}


%

\end{document}